\documentclass[letterpaper, 12 pt, conference]{article} 
\usepackage[letterpaper, portrait, margin=1in]{geometry} 
\usepackage{newtxtext,newtxmath} 
\usepackage[font=small]{caption} 
\usepackage{authblk} 
\usepackage{indentfirst} 
\usepackage{graphics}
\usepackage{epsfig}
\usepackage{amsmath}
\usepackage{rotating} 
\usepackage{wrapfig}
\usepackage{lscape}
\usepackage{amssymb}
\usepackage{amsfonts}
\usepackage{cite}
\usepackage{algorithmic}
\usepackage{algorithm}
\usepackage{psfrag}
\usepackage{enumerate}
\usepackage{array,multirow,longtable} 
\usepackage{color}
\usepackage{booktabs} 
\usepackage{soul}
\usepackage{accents} 
\usepackage[title]{appendix}
\usepackage{etoolbox, apptools} 
\AtBeginEnvironment{appendices}{\appendixtrue}
\renewcommand{\baselinestretch}{1}
\makeatletter 
\patchcmd{\@maketitle}{\LARGE \@title}{\fontsize{12}{12.5}\selectfont\@title}{}{}
\makeatother

\usepackage{titlesec} 
\titleformat{\section}[block]{\color{black}\fontsize{12}{15}\bfseries\filcenter}{\thesection}{1em}{}
\titleformat{\subsection}[block]{\color{black}\fontsize{12}{15}\bfseries}{\thesubsection}{1em}{}
\titleformat{\section}[block] 
  {\normalfont\bfseries\filcenter}{\IfAppendix{\appendixname~}{\relax}\thesection\IfAppendix{: }{}}{1em}{}

\titleformat{\subsection}
  {\normalfont\bfseries}{\thesubsection}{1em}{}

\title{\bf TEMPERATURE STATES IN POWDER BED FUSION ADDITIVE MANUFACTURING ARE STRUCTURALLY CONTROLLABLE AND OBSERVABLE}

\author[1]{Nathaniel Wood}
\author[1]{David J. Hoelzle \thanks{Corresponding author. Phone: +1 (614) 688-2942; email: hoelzle.1@osu.edu}}
\affil[1]{Department of Mechanical and Aerospace Engineering, the Ohio State University, Columbus, OH 43210}

\date{ }

\newtheorem{thm}{Theorem}

\newtheorem{defn}{Definition}
\newtheorem{rem}{Remark}

\begin{document}
\maketitle
\vspace{-.2in}
\section*{Abstract}
Powder Bed Fusion (PBF) is a type of Additive Manufacturing (AM) technology that builds parts in a layer-by-layer fashion out of a bed of metal powder via the selective melting action of a laser or electron beam heat source.  The technology has become widespread, however the demand is growing for closed loop process monitoring and control in PBF systems to replace the open loop architectures that exist today. Controls-based models have potential to satisfy this demand by utilizing computationally tractable, simplified models while also decreasing the error associated with these models.  This paper introduces a controls theoretic analysis of the PBF process, demonstrating models of PBF that are asymptotically stable, stabilizable, and detectable.  We show that linear models of PBF are structurally controllable and structurally observable, provided that any portion of the build is exposed to the energy source and measurement, we provide conditions for which time-invariant PBF models are classically controllable/observable, and we demonstrate energy requirements for performing state estimation and control for time-invariant systems. This paper therefore presents the foundation for an effective means of realizing closed loop PBF quality control.
\section{Introduction}\label{sec:In}

\begin{figure*}[!tb]
\centering
\includegraphics[width=0.89\linewidth]{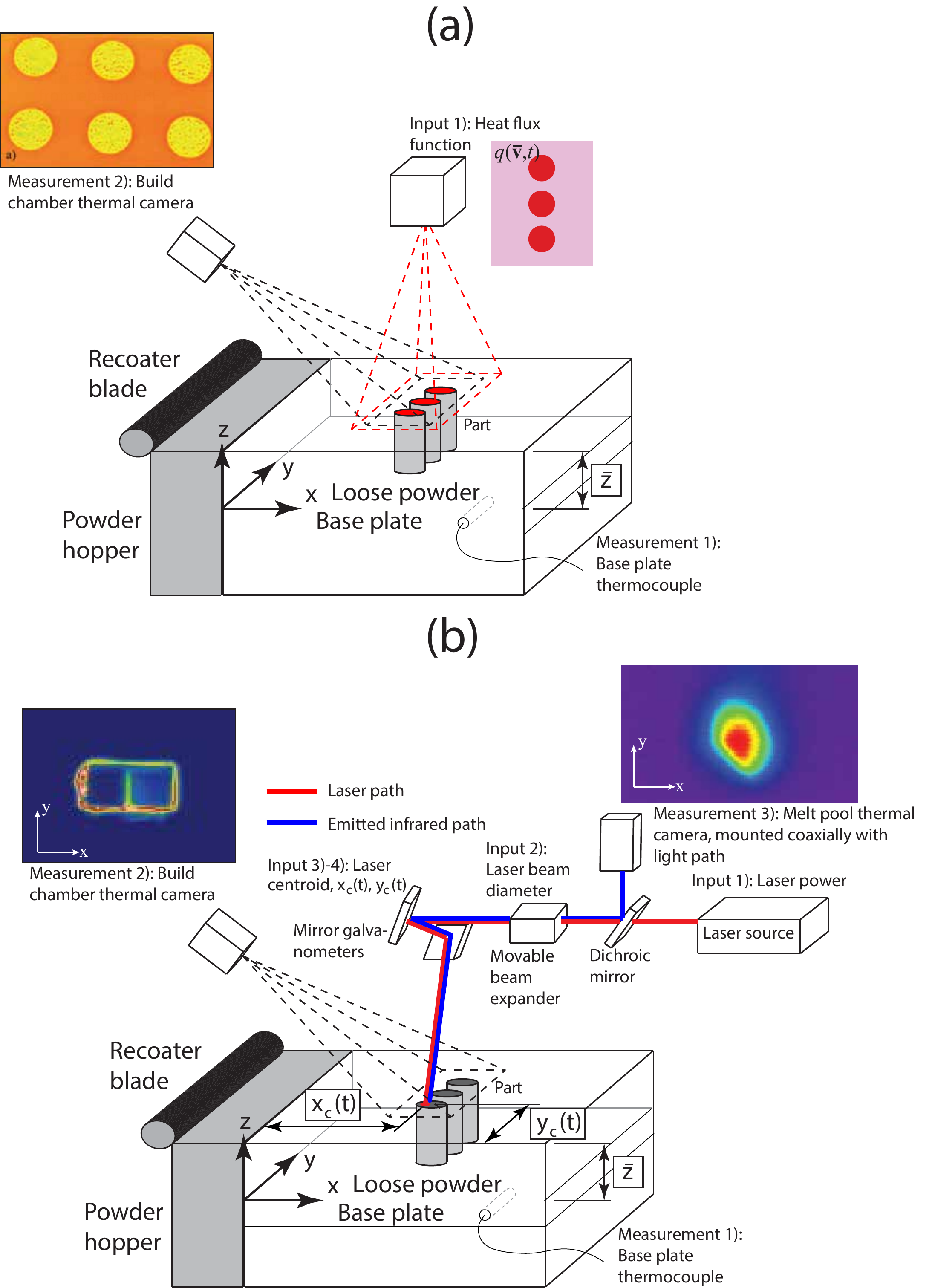}
\caption{\textbf{System schematic of Powder bed fusion (PBF) additive manufacturing.} a) Input and output channels for E-PBF and DLP-PBF.  Measurement (2) screenshot reproduced from \cite{Ridwan_2014}. b) Input and output channels for L-PBF.}
\label{fig:PBF}
\end{figure*}

 \emph{Powder Bed Fusion} (PBF) belongs to a class of manufacturing processes known as \emph{additive manufacturing} (AM).  Commonly referred to as ``three-dimensional (3-D) printing,'' these processes have rapidly grown in popularity and market size due to their ability to produce parts of complex geometry, with engineering properties meeting or exceeding those produced by conventional manufacturing processes, while removing the majority of the overhead costs normally associated with production \cite{Bhavar_2014,Wang_2015_grains,Brandl_2011}. The PBF process (Fig. \ref{fig:PBF}) builds three-dimensional parts out of layers of metal powder, using a build cycle consisting of three stages: 1) sweeping a thin layer of powder over a base of metal feedstock or previously-applied powder, 2) selectively melting a pattern of desired geometry into the powder by application of a high-powered laser or electron beam (e-beam), and 3) lowering the build platform in the $-z$ direction to accommodate a fresh layer of powder.  

\par The PBF process is not without flaws.  It is well-documented that parts manufactured with PBF display high levels of residual stresses \cite{Peng_2016a_article,Peng_2016b_article, krol_2013}, porosity \cite{Gokuldoss_2017,DebRoy_2017,Mower_2016}, and anisotropy in material properties \cite{Keist_2016,Wang_2015_grains,Wei_2016,Mower_2016,Yadollahi_2015,Parimi_2014}, and that these defects are a direct consequence of the thermal management of the PBF process during production.  Although thermal management is critical for the manufacture of high-quality parts, current PBF machines operate in open-loop, with the irradiated energy to the system, $u$, specified by a schedule directed by G-Code machine language \cite{Yeung_2018}.  Appropriate parameter values that govern the schedule are determined through operator experience, heuristically through design-of-experiment procedures \cite{Arisoy_2017} and/or with computationally complex predictive models \cite{Khairallah_2016}.  
Significant advances in PBF production quality could be achieved with feedback control of the thermal management problem.  Despite the need for thermal management of PBF, the community has not established the theory to evaluate the basic criteria for modern control synthesis: the requirement that the process is controllable and observable.  This paper answers this basic question.  Our analysis is aspirational, considering both current and emerging thermal actuation and sensing hardware capabilities and we do not consider computational constraints during model construction.  The aim is to establish a controls theoretic basis for PBF, thus providing a framework to apply modern controls tools such as multivariable robust controllers, state estimators, and fault detection schemes to this important, emerging manufacturing modality.  

Throughout this paper we reference the nomenclature tabulated in Table \ref{tb: Nomenclature}.

\begin{longtable}[h]{cc}
\caption{Nomenclature used throughout paper}\\
\hline\hline
{Signal} & {value}\\
\hline 
{$q(t)$} & {Heat flux}\\
{$\mathbf{x}(t)$} & {State signal (continuous)}\\
{$\mathbf{x}[k]$} & {State signal (discrete)}\\
{$\mathbf{u}(t)$} & {Input signal (continuous)}\\
{$\mathbf{u}[k]$} & {Input signal (discrete)}\\
{$\mathbf{y}(t)$} & {Output signal (continuous)}\\
{$\mathbf{y}[k]$} & {Output signal (discrete)}\\\cline{1-2}
{Classifier} & {Description}\\\cline{1-2}
{$m,n,p,r$} & {Matrix size variables}\\
{$\mathbb{R}^n$} & {Set of all real-valued $n$-dimensional vectors}\\
{$\mathbb{R}^{m\times n}$} & {Set of all real-valued matrices of size $m\times n$}\\
{$\mathcal{X}$} & {Set of all matrices $\mathbf{X}$ with constant pattern}\\
{$||\mathbf{X}||$, $\mathbf{X}\in\mathbb{R}^{m\times n}$} & {Generic norm of a matrix}\\
{$|\mathcal{Y}|$} & {Cardinality of a set $\mathcal{Y}$}\\
{$\cdot$} & {Vector dot product}\\
{$\epsilon$} & {Small positive constant}\\
{$i,j,s,l$} & {Generic indexing variables}\\
{$e$} & Indexing over elements in finite element mesh\\
{$\mathbf{X}>0$, $\mathbf{X}\in\mathbb{R}^{n\times n}$} & {$\mathbf{X}$ is positive definite}\\
{$[\mathbf{X}]_{ij}$, $\mathbf{X}\in\mathbb{R}^{m\times n}$} & {Selects the $(ij)^{th}$ element of $\mathbf{X}$}\\\cline{1-2}
{Symbol} & {Description}\\\cline{1-2}
{$\mathbf{v}=[x,y,z]^{\prime}$} & {Spatial coordinate}\\
{t} & {Time coordinate}\\
{$V$} & {Build domain}\\
{$S$} & {Boundary of build domain}\\
{$\Omega$} & {Top surface of build domain}\\
{$\Lambda$} & {Bottom surface of build domain}\\
{$\Gamma$} & {All other surfaces of build domain}\\
{$T$} & {Temperature}\\
{$c$} & {Specific heat}\\
{$\rho$} & {Density}\\
{$k$} & {Thermal conductivity}\\
{$\mathbf{\kappa}$} & {Directional thermal conductivity}\\
{$P$} & {Heat source power}\\
{$\sigma^2$} & {Heat source variance}\\
{$Bi$} & {Biot number}\\
{$\nabla$} & {Del operator}\\
{$\Pi$} & {Functional form of partial differential equation}\\
{$\mathbf{M}$} & {FEM heat capacitance matrix}\\
{$\mathbf{K}$} & {FEM thermal conductivity matrix}\\
{$\mathbf{R}(t)$} & {FEM load vector}\\
{$\mathbf{N}_e$} & {FEM shape functions}\\
{$\mathbf{B}_e$} & {FEM shape function gradient}\\
{$\mathbf{\Phi}(t,t_0)$} & {State transition matrix}\\
{$W_c(t_0,t_1)$} & {Controllability gramian}\\
{$W_o(t_0,t_1)$} & {Observability gramian}\\
{$\mathbf{A}(t)$} & {State dynamics matrix}\\
{$\mathbf{B}(t)$} & {Input-to-state mapping matrix}\\
{$\mathbf{C}(t)$} & {State-to-output mapping matrix}\\
{$\mathcal{C}$} & {Controllability matrix}\\
{$\mathcal{O}$} & {Observability matrix}\\
{$G(\mathbf{A,B,C})$} & {Graph of linear system}\\
{$N$} & {Set of nodes of $G(\mathbf{A,B,C})$}\\
{$\lambda_i$} & {Eigenvalue of a matrix}\\
{$\mathbf{V}_i$} & {Eigenvector associated with $\lambda_i$}\\
{$\delta(\lambda_i)$} & {Algebraic multiplicity of $\lambda_i$}\\
{$\mu(\lambda_i)$} & {Geometric multiplicity of $\lambda_i$}\\
\hline\hline
\label{tb: Nomenclature}
\end{longtable}

\subsection{PBF actuation and sensing structure}\label{sec:Intro_thermal_structure}
\begin{figure} [!tb]
\centering
\includegraphics[width=0.65\textwidth]{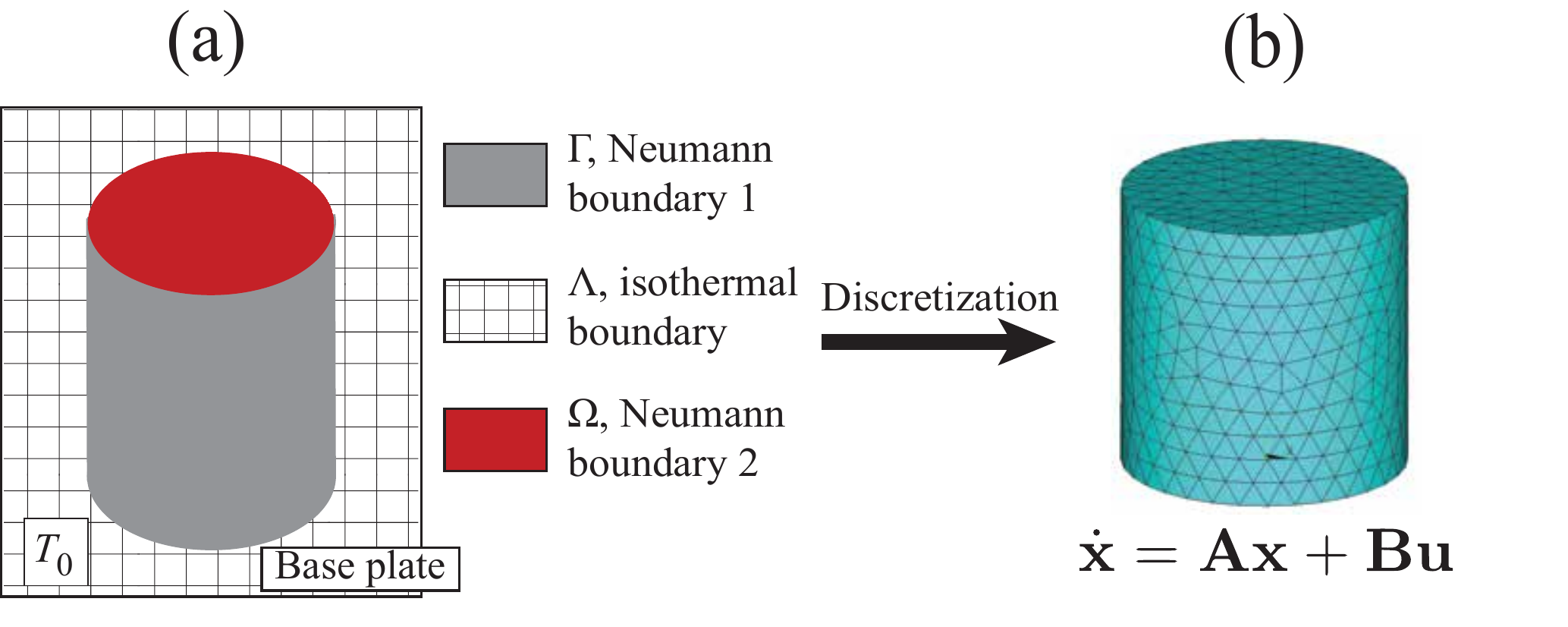}
\caption{\textbf{Transition from heat conduction model of PBF to Finite Element Method (FEM) description of system.}  (a) Description of PBF boundary conditions for a simple build layer.  (b) FEM-based discretization of PBF solution domain and associated system of coupled ODEs.}
\label{fig: PBF_BCs}
\end{figure}

Consider a partially built part in a PBF system (Fig. \ref{fig:PBF}).  The part is the thermal domain, $V$, with heat transfer defined on the domain $\mathbf{v}=\{x,y,z\}\in V\subset\mathbb{R}^3$ (Fig. \ref{fig: PBF_BCs}a).  This domain is bounded by the set of faces $S=\{\Lambda,\Gamma,\Omega\}\subset\mathbb{R}^3$, defined below:

\begin{itemize}
\item $\Lambda$ contains all faces at the ``bottom'' of the part, consisting of points $\mathbf{\underaccent{\bar}{v}}=\{x,y,0\}\in V$, which are in contact with the machine base plate. 
\item $\Omega$ contains all faces at the ``top'' of the part, consisting of points $\mathbf{\bar{v}}=\{x,y,\bar{z}\}\in V$, which are exposed to the environment, laser or e-beam energy sources, and vision-based thermal sensors. 
\item $\Gamma$ contains all other bounding faces of the part, which are in contact with the surrounding metal powder.  
\end{itemize}

Temperatures within $V$, $T(\mathbf{v},t)$, $\mathbf{v}\in V$, time $t\in\mathbb{R}_{+}$, are defined by the resultant heat flux balance

\begin{equation}
C\frac{dT(\mathbf{v},t)}{dt}=u-q_{\text{cond.}}-q_{\text{conv}}-q_{\text{rad}}.\label{eq: FluxBalance}
\end{equation}

\noindent where $u(\mathbf{\bar{v}},t)$ is the thermal energy delivered to the top face $\mathbf{\bar{v}}\in\Omega$ from either laser or e-beam irradiation, $q_{\text{cond}}$ describes heat transfer via conduction, $q_{\text{conv}}$ describes heat transfer via convection, $q_{\text{rad}}$ describes heat transfer via radiation, and $C=\rho c$ describes the heat capacitance of a material with density $\rho$ and specific heat capacity $c$.  As Fig. \ref{fig: PBF_BCs} demonstrates, we construct PBF models based on approximating the flux balance \eqref{eq: FluxBalance} with the Finite Element Method (FEM). We are therefore interested in a set of temperaures in $V$ at the specific spatial locations of the $n$ nodes in our FEM mesh, $\mathbf{v}_1, \mathbf{v}_1, \ldots, \mathbf{v}_n$, which we will term the state vector $\mathbf{x}(t)\in\mathbb{R}^n$,  

\begin{equation}
    \mathbf{x}(t)=\left[T(\mathbf{v}_1,t),T(\mathbf{v}_2,t),\ldots, T(\mathbf{v}_n,t)\right]^\prime.
    \label{eq:state_disc}
\end{equation}

\noindent $\left[\cdot\right]^\prime$ denotes the transpose operator.

\subsubsection{Actuation (A) Modes}\label{sec:intro_actuation}
PBF systems are currently actuated, or proposed to be actuated, via three different modes.  Each of these modes governs the structure of $u(\bar{\mathbf{v}},t)$ incident on the build.  

\begin{enumerate}
\item[\textbf{A.A)}] \textbf{$u(\mathbf{\bar{v}},t)$ as an arbitrary function.}  Heat flux function $u(\mathbf{\bar{v}},t)$ can be arbitrarily applied (Fig. \ref{fig:PBF}a).  Actuation mode \textbf{A.A} is applicable to an e-beam system (E-PBF) operating in a mode which is capable of bringing large portions of $\Omega$ to a common temperature simultaneously \cite{Rafi_2013,Antonysamy_2013}.  This is possible because the electron beam is capable of rastering with speeds reaching $\sim10^4$ mm/s \cite{Ramirez_2011}, which meets the critical raster speed for heat input across successive rasters to outpace heat loss due to conduction as demonstrated in \cite{Juechter_2014}.  A laser PBF (L-PBF) system that uses a digital light processing filter to distribute radiant energy as a function (DLP-PBF) is also capable of producing arbitrary heat flux functions across $\Omega$ \cite{Roy_2018}.   

\item[\textbf{A.B)}] \textbf{$u(\mathbf{\bar{v}},t)$ as a Gaussian function with control channels 1 -- 4.}  Laser or e-beam incident energy is assumed to have a Gaussian-distributed intensity \cite{Raghavan_2013,Yuan_2015} and a total of $M$ heat sources (spots) are assumed active at any time $t$:

\begin{equation}
u(\mathbf{\bar{v}},t)=\sum_{i=1}^M\frac{P_i}{\sqrt{2\pi\sigma_i^2}}\text{exp}\left({-\frac{1}{2}\frac{(x_{c,i}-x_{\bar{\mathbf{v}}})^2+(y_{c,i}-y_{\bar{\mathbf{v}}})^2}{\sigma_i^2}}\right).\label{eq: HeatSource}
\end{equation}

\noindent The system provides complete control authority over control channels 1 -- 4 in Fig. \ref{fig:PBF}: 1) Laser or e-beam effective power, $P_i(t)$; 2) Laser or e-beam variance, $\sigma_i^2(t)$; and 3 \& 4) Laser or e-beam centroid, $x_{c,i}(t)$ and $y_{c,i}(t)$.  Actuation model \textbf{A.B} is applicable to multi- ($M>1$) and single- ($M=1$) laser L-PBF and E-PBF operating in a mode in which the raster speed is on the order of the effective time constant of the thermal system.  

\item[\textbf{A.C)}] \textbf{$u(\mathbf{\bar{v}},t)$ as a Gaussian function with control channels 1 -- 2.}  \textbf{A.C} represents the minimal control fidelity currently envisioned and used in practice.  The Gaussian laser or e-beam centroids of \eqref{eq: HeatSource} are not control variables; instead the centroids $x_{c,i}(t)$ and $y_{c,i}(t)$ proceed on a schedule as dictated by G-Code commands.
\end{enumerate}

\subsubsection{Measurement (M) Modes} \label{sec:intro_measurement}
PBF system temperature is currently measured, or proposed to be measured, via three different modes, or combinations thereof.   

\begin{enumerate}
    \item[\textbf{M.1)}]\textbf{Environmental temperature measurement, $T_0$, via a thermocouple embedded in the baseplate or surrounding system.}  This is the most basic temperature measurement available.  This sensor is used for real-time feedback of environmental temperature and is commonly available in commercial PBF systems.  We model this measurement as $y=h(\bar{x},\bar{y},\epsilon,\epsilon)T(x,y,z=0,t)$, where $h$ represents a generic windowing function with square window $\epsilon$ \footnotesize$\gtrapprox\:$\normalsize  0 that is centered at the point $(\bar{x},\bar{y},z=0)$.  $\epsilon$ represents the element size of a single point thermocouple. 
    \item[\textbf{M.2)}]\textbf{Fixed field of view thermal camera.}  The camera has a fixed field of view (FOV) of the face $\Omega$ to collect emitted infrared light from the part \cite{rodriquez_2012}, $y=h(\bar x, \bar y, M,N)T(\bar{\mathbf{v}},t)$, where $h$ is a generic $M$ by $N$ windowing function centered at the point $\left(\bar x,\bar y,z=\bar z\right)$.  In cases where the measurement is a single point (spot) pyrometer centered at $\left(\bar x,\bar y\right)$ \cite{Cola_2018}, the windowing function is given by $h(\bar x, \bar y, \epsilon, \epsilon)$, where $\epsilon$ \footnotesize$\gtrapprox\:$\normalsize  0 is the element size of the single point pyrometer.  \textbf{M.2} is possible for both E-PBF and L-PBF systems.
    \item[\textbf{M.3)}]\textbf{Source-centered field of view thermal camera.}  Emitted infrared light from face $\Omega$ passes through the mirror galvonometers (Fig. \ref{fig:PBF}b) before being split off to a single point pyrometer \cite{Chivel_2010}, pyrometer array \cite{Stockman_2018}, and/or infrared (IR) camera \cite{Clijsters_2014}.  As the FOV is coaxial with the laser centroids  $y=\left(\sum_{i=1}^Mh(x_{c,i}+\bar x,y_{c,i}+\bar y,M_i,N_i)\right)T(\bar{\mathbf{v}},t)$; $\bar x$ and $\bar y$ capture an offset between the laser centroid and the center of a camera FOV.  In cases where the measurement is a single point pyrometer, the windowing function is given by $h\left(x_{c,i}+\bar x,y_{c,i}+\bar y, \epsilon,\epsilon\right)$.  \textbf{M.3} is only possible with L-PBF.  
\end{enumerate}




\subsection{Contributions of the paper} The main contribution of this manuscript is the definition of a systems-based analysis of controllability and observability of the PBF process, which sets the stage for rigorous controller and estimator analysis and design.  The intended audiences are additive manufacturing researchers who are interested in process control and control systems researchers who are interested in the PBF application.  As such, we provide basic control theory definitions and systems descriptions and state these definitions in the context of PBF.  Central to this work are the definitions of controllability and observability.

\begin{defn}{Controllability.}\label{def:control}
We use the standard definition for controllability given by \cite{chen_1999}: a system is ``said to be controllable if for any initial state $\mathbf{x}(t_0)=\mathbf{x}_0$ and any final state $\mathbf{x}_f$, there exists an input that transfers $\mathbf{x}_0$ to $\mathbf{x}_f$ in finite time.  Otherwise the system is said to be uncontrollable.'' 
\end{defn}

\begin{defn}{Observability.}\label{def:observe}
We use the standard definition for observability given by \cite{chen_1999}: a system is ``said to be observable if for any initial state $\mathbf{x}(t_0)=\mathbf{x}_0$, there exists a finite $t_f>0$ such that knowledge of the input $
\mathbf{u}$ and the output $\mathbf{y}$ over $\left[t_0, t_f\right]$ suffices to determine uniquely the initial state $\mathbf{x}_0$. Otherwise the system is said to be unobservable.''
\end{defn}

In the context of PBF, Definition \ref{def:control} states that PBF is controllable if there always exists an actuation function $u(\bar{\mathbf{v}},t)$ such that the set of temperature states in the part, $\mathbf{x}(t)$, can be driven from a set of initial temperature values to any set of final temperature values.  In practice, this means that the temperature field inside the part can be sculpted in time.  In the context of PBF, Definition \ref{def:observe} states that PBF is observable if we can estimate the history of temperature states from knowledge of the actuation function $u(\bar{\mathbf{v}},t)$ and the measurement function $\mathbf{y}(t)$.  In practice, this means that we can estimate the temperature field inside the entire part domain from knowledge of the actuation function and measurement readings.  

To communicate with the intended audience, this paper translates established control and network theory for system models of PBF.  Section \ref{sec:Preliminaries} surveys the established theoretical results in systems and network theory that are leveraged in this manuscript, namely structural controllability and observability.  Section \ref{sec:thermal_model} expands on the introduction of the thermal model in Section \ref{sec:Intro_thermal_structure} to detail simplifying assumptions and the methodology for spatially discretizing the partial differential equation (PDE) in \eqref{eq: FluxBalance} into a network of ordinary differential equations (ODEs).  Sections \ref{sec: cont_obs_case_1}-\ref{sec: cont_obs_cases_2_4} prove conditions under which the different actuation and measurement modes of PBF are controllable and observable.  Section \ref{sec: control energy} establishes a more practical evaluation of controllability and observability, providing ways to measure the energy required to drive certain temperature states or estimate certain temperature states.  Section \ref{sec:Con} postulates a set of future research directions that these controllabilty and observability properties of PBF allow the control engineer to pursue.

\section{Preliminaries}\label{sec:Preliminaries}
Throughout this paper we will denote scalar variables and functions by italicized variables and vector-valued functions as bold-face variables: for example $T$ is the 3-D part temperature function whereas $\mathbf{x}$ is a vector of discretized temperature nodes.  $\mathbf{I}_m$ is the $m\times m$ identity matrix, $\mathbf{0}_{m}$ is the $m\times m$ zero matrix.  For a generic matrix $\mathbf{P}$, $\mathbf{P}>0$ denotes that $\mathbf{P}$ is positive definite (PD) \cite{chen_1999}.  The $ij^{th}$ entry of a matrix $\mathbf{M}$ is denoted as $[\mathbf{M}]_{ij}$.

We shall invoke concepts of controllability, observability, \emph{structural controllability} and \emph{structural observability} for dynamic systems of the form

\begin{equation}
    \begin{aligned}
    \dot{\mathbf{x}} &= \mathbf{f}(t,\mathbf{x})+\mathbf{r}(t,u)\\
    \mathbf{y} &= \mathbf{g}(t,\mathbf{x}).\\
    \end{aligned}
    \label{eq:nonlinear_generic}
\end{equation}

\noindent $\mathbf{f}$, $\mathbf{r}$, and $\mathbf{g}$ are nonlinear, vector-valued functions of the system.  Analogous to the discretization used to construct the state vector $\mathbf{x}$ in \eqref{eq:state_disc}, our analysis will investigate discretizations of the input and output functions: $\mathbf{u}(t)\in\mathbb{R}^m$ is the input vector, and $\mathbf{y}(t)\in\mathbb{R}^p$ is the output vector. For certain actuation and measurement modes, \eqref{eq:nonlinear_generic} will simplify to either a linear time-varying (LTV) system

\begin{equation}
    \begin{split}
        &\dot{\mathbf{x}}=\mathbf{A}\mathbf{x}+\mathbf{B}(t)\mathbf{u}\\
        &\mathbf{y}=\mathbf{C}(t)\mathbf{x},
    \end{split}\label{eq: Sample_LTV_sys}
\end{equation}

\noindent With $\mathbf{A}(t)\in\mathbb{R}^{n\times n}$, $\mathbf{B}(t)\in\mathbb{R}^{n\times p}$, and $\mathbf{C}(t)\in\mathbb{R}^{q\times n}$, or the linear time-invariant (LTI) system 

\begin{equation}
    \begin{split}
        &\dot{\mathbf{x}}=\mathbf{A}\mathbf{x}+\mathbf{B}\mathbf{u}\\
        &\mathbf{y}=\mathbf{C}\mathbf{x}.
    \end{split}\label{eq: Sample_LTI_sys}
\end{equation} 

\subsection{Controllability and Observability criteria for LTV and LTI systems}\label{subsec: classial_cont_obs}

The following are standard textbook definitions from \cite{chen_1999} and \cite{Antasaklis_2007}; the intention is to develop a set of preliminaries to reference throughout the paper.  The criteria for controllability and observability is built from the definition of the state transition matrix, $\mathbf{\Phi}(t,t_0)$, which describes the mapping from initial state $\mathbf{x}_0$ and input function $\mathbf{u}(t)$ to the state at time $t$, 

\begin{equation}
    \mathbf{x}(t)=\mathbf{\Phi}(t,t_0)\mathbf{x}_0+\int_{t_0}^t\mathbf{\Phi}(t,\tau)\mathbf{B}(\tau)\mathbf{u}(\tau)d\tau.
    \label{eq:LTV_state_trans}
\end{equation}

\noindent An input $\mathbf{u}(t)$ can be constructed to drive the system from state $\mathbf{x}(0)$ to any $\mathbf{x}(t)$, thus the system is controllable, if and only if the controllability grammian

\begin{equation}\label{eq: LTV_Gram_Cont}
    W_{c}(t_0,t_1)=\int_{t_0}^{t_f}\mathbf{\Phi}^\prime(t_1,\tau)\mathbf{B}(t)\mathbf{B}^\prime(t)\mathbf{\Phi}(t_1,\tau)d\tau,
\end{equation}

\noindent is non-singular.  Likewise, any initial state $\mathbf{x}_0$ can be reconstructed from knowledge of the input $\mathbf{u}$ and output $\mathbf{y}$, thus the system is observable, if and only if the observability grammian

\begin{equation}
W_o(t_0,t_1)=\int_{t_0}^{t_f}\mathbf{\Phi}^\prime(t_1,\tau)\mathbf{C}^\prime(\tau)\mathbf{C}(\tau)\mathbf{\Phi}(t_1,\tau)d\tau,
\label{eq:LTV_cont}
\end{equation}

\noindent is non-singular.  In general, $\mathbf{\Phi}(t,t_0)$ is difficult to compute.  For the specific case where the $\mathbf{A}$-matrix in \eqref{eq: Sample_LTI_sys} is time invariant, $\mathbf{A}(t)=\mathbf{A}$, the state transition matrix simplies to $\mathbf{\Phi}(t,t_0)=e^{\mathbf{A}(t-t_0)}$.  Furthermore, in cases where the system is LTI, \eqref{eq: Sample_LTI_sys}, the criteria for controllabilty and observability can be simplified to a matrix rank test.  As stated in \cite{Antasaklis_2007}, the pair $\left(\mathbf{A},\mathbf{B} \right)$ is controllable if controllability matrix

\begin{equation}
\mathcal{C}=\begin{bmatrix}
\mathbf{B} & \mathbf{AB} & \cdots & \mathbf{A}^{n-1}\mathbf{B},
\end{bmatrix},
\end{equation}

\noindent has rank $n$.  Likewise, the pair $\left(\mathbf{A}, \mathbf{C}\right)$ is observable if the observability matrix:

\begin{equation}
\mathcal{O}=\begin{bmatrix}\mathbf{C}\\\mathbf{CA}\\\vdots\\\mathbf{CA}^{n-1}\end{bmatrix},\end{equation}

\noindent has rank $n$.  

\subsection{Structural Controllability and Observability}\label{subsec: struc_cont_obs}
In general, the number of temperature states $n$ is large in PBF (order of $10^3$ -- $10^4$ states \cite{Wood_2018}). Accordingly, rank tests of $\mathcal{C}$ and $\mathcal{O}$ are inefficient and sensitive to parameter variation for large $n$ \cite{paige_1981}.  The notions of structural controllability (SC) and  structural observability (SO) abstract the ideas of controllability/observability away from the particular construction of $(\mathbf{A,B})$ and $(\mathbf{A,C})$, and instead presents them as inherent properties of the given network topology.  This is a powerful tool for assessing system controllability/observability in the face of uncertain model parameters.  For example, for the same system two different parameter estimations may result in the construction of pairs $(\mathbf{A_0,B_0})$ and $(\mathbf{A_1,B_1})$ in which the former is uncontrollable and the latter is controllable, despite the network topology (system structure) remaining constant.  Our treatment of this topic follows the presentation given in \cite{Liu_YY_2011} and is meant to be a brief introduction to the field; readers interested in a more complete description of the subject should consult \cite{Liu_YY_2011}.\par

\begin{figure*} [!tb]
\centering
\includegraphics[width=1.0\textwidth]{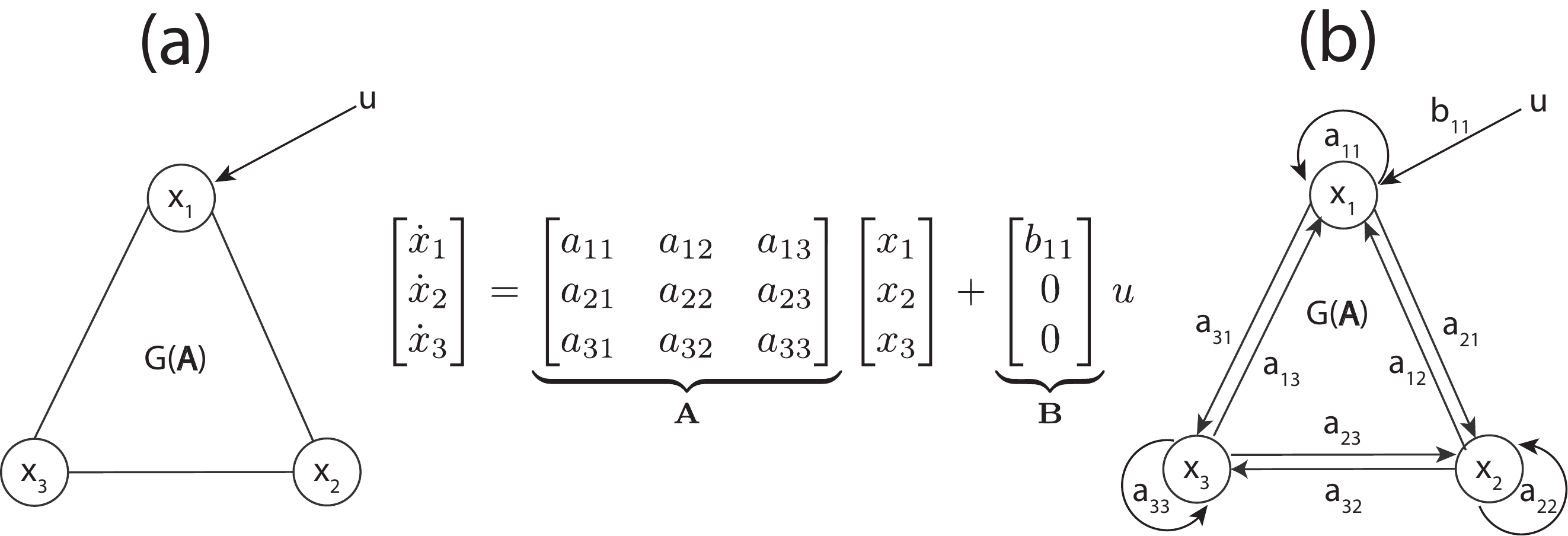}
\caption{\textbf{Interpretation of LTI system as a graph.}  (a) Simple model consisting of one element and three nodes, with external input being applied to node 1.  $\mathbf{A}$ matrix maps conductive heat flow between nodes with all diagonal elements being nonzero. $\mathbf{B}$ matrix maps the input $u$ to the system dynamics.  (b) Interpretation of system as a directed network following notation of \cite{Liu_YY_2011}, with links (edges) of network mapping heat flow between nodes according to the weights specified by $\mathbf{A}$ and $\mathbf{B}$.  Self-loops within network correspond to energy storage within nodes.}
\label{fig: StructuralControllability}
\end{figure*}

We present the theory of SC and SO by constructing a \emph{graph} $G(\mathbf{A})$ from $\mathbf{A}$ of \eqref{eq: Sample_LTI_sys}:  $G(\mathbf{A})$ is a collection of nodes and edges.  The nodes of $G(\mathbf{A})$ are defined as the state components of $\mathbf{x}$.  Each nonzero entry $[\mathbf{A}]_{ij}$ of $\mathbf{A}$ corresponds to an edge of $G(\mathbf{A})$ directed from node $j$ to node $i$ and having a link weight equal to $[\mathbf{A}]_{ij}$.  The input $\mathbf{u}$ is passed into $G(\mathbf{A})$ through a set of \emph{driver nodes}.  The component $\mathbf{u}_j$ of $\mathbf{u}$ is connected to its driver nodes by a set of edges which are defined and weighted by the nonzero elements $[\mathbf{B}]_{ij}$ of the $j^{th}$ column of $\mathbf{B}$.   Fig. \ref{fig: StructuralControllability} demonstrates this interpretation on an extremely simple graph representation of a discretized thermal domain.  Similarly, the output $\mathbf{y}$ is received from $G(\mathbf{A})$ through a set of \emph{observer nodes}.  The component $\mathbf{y}_i$ of $\mathbf{y}$ is connected to its observer nodes by a set of edges which are defined and weighted by the nonzero elements $[\mathbf{C}]_{ij}$ of the $i^{th}$ row of $\mathbf{C}$.  $G(\mathbf{A})$ therefore is a graph whose edges represent the flow of state information from the input to the system states, between system states, and from system states to the output. The matrices $\mathbf{B}$ and $\mathbf{C}$ ``select'' the driver and observer nodes of the system.   

We construct the graph $G(\mathbf{A},\mathbf{B})$ by treating each component of $\mathbf{u}$ as a node (an ``input node'') and appending these nodes to the set of nodes that define $G(\mathbf{A})$.  The edges between each node $\mathbf{u}_j$ and its corresponding driver nodes, which construct $\mathbf{B}$, are appended to the set of edges that define $G(\mathbf{A})$.  The graph $G(\mathbf{A},\mathbf{C})$ is constructed by treating the output $\mathbf{y}$ (the ``output nodes'') and its corresponding observer nodes similarly.  The graph $G(\mathbf{A},\mathbf{B},\mathbf{C})$ may be constructed by appending both $\mathbf{u}$ and $\mathbf{y}$ to $G(\mathbf{A})$.

Section \ref{sec: cont_obs_case_1}-\ref{sec: cont_obs_cases_2_4} will show how the thermal models \eqref{eq:nonlinear_generic}-\eqref{eq: Sample_LTI_sys} relate to $G(\mathbf{A})$.  Others have developed different graph theoretic representations of PBF \cite{Yavari_2019}.  

Analysis of SC/SO of $G(\mathbf{A})$ assumes that the nonzero values of $\mathbf{A}$, $\mathbf{B}$, and $\mathbf{C}$ are uncertain, which is true for any model produced from a system with imprecisely-known parameters.  We supply the definition of SC as given in \cite{Liu_YY_2011} below.

\begin{defn}
\eqref{eq: Sample_LTI_sys} is SC if the uncertain nonzero (free) parameters (entries) of $\mathbf{A}$ and $\mathbf{B}$ may be chosen such that the resultant system $(\mathbf{A,B})$ is controllable in the sense of Section \ref{subsec: classial_cont_obs}. \label{def: structural_controllability} 
\end{defn} 

The problem of SO of $G(\mathbf{A})$ is equivalent to the problem of SC of $G(\mathbf{A}^\prime)$ \cite{Liu_YY_2013}.\par 

Furthermore, as \cite{Liu_YY_2011} states, controllability/observability is a \emph{dense} set.  In the case of controllability, this property means that if a certain model parameter estimation results in an uncontrollable pair $(\mathbf{A_0,B_0})$ then there exists a completely controllable pair $(\mathbf{A_1,B_1})$ resulting from a different estimation such that $\lVert\mathbf{A_1-A_0}\rVert<\epsilon$ and $\lVert\mathbf{B_1-B_0}\rVert<\epsilon$ for all $\epsilon>0$.  An infinitesimally small error in the estimated parameter values may produce a system $(\mathbf{A_1,B_1})$ that is controllable despite predictions of uncontrollability.  In posing the question of controllability as an inherent property of the model structure, rather than a consequence of specific parameter estimations, SC removes ambiguity due to parameter and numerical uncertainty. An analogous argument can be made for observability \cite{Liu_YY_2013}.\par

The concepts of SC/SO may be extended to time-varying systems \eqref{eq: Sample_LTV_sys}.  We use the extension given by \cite{Reissig_2014,Hartung_2013_Necessary,Hartung_2013_Sufficient,Hartung_2012}, which is summarized below:

\begin{defn}
Let $\mathbf{A}(t)\in\mathbb{M}^{n\times n}$, $\mathbf{B}(t)\in\mathbb{M}^{n\times m}$, $\mathbf{C}(t)\in\mathbb{M}^{p\times n}$.  Define the following sets over a given time interval $\mathbb{T}=[t_0,t_1]$:
\begin{itemize}
    \item $\mathcal{A}\coloneqq\left\{\mathbf{X}\in\mathbb{R}^{n\times n}:\: [\mathbf{X}]_{ij}\neq0\Leftrightarrow[\mathbf{A}]_{ij}(t)\neq0\:\forall\:t\in\mathbb{T}\right\}$
    \item $\mathcal{B}\coloneqq\left\{\mathbf{X}\in\mathbb{R}^{n\times m}:\: [\mathbf{X}]_{ij}\neq0\Leftrightarrow[\mathbf{B}]_{ij}(t)\neq0\:\forall\:t\in\mathbb{T}\right\}$
    \item $\bar{\mathcal{C}}\coloneqq\left\{\mathbf{X}\in\mathbb{R}^{p\times n}:\: [\mathbf{X}]_{ij}\neq0\Leftrightarrow[\mathbf{C}]_{ij}(t)\neq0\:\forall\:t\in\mathbb{T}\right\}$
\end{itemize}
These sets characterize the \textit{patterns} of $\mathbf{A}$, $\mathbf{B}$ and $\mathbf{C}$.  LTV systems \eqref{eq: Sample_LTV_sys} with constant patterns are said to be SC on $\mathbb{T}$ if there exists $\mathbf{A}(t)\in\mathcal{A}$ and $\mathbf{B}(t)\in\mathcal{B}$ such that $(\mathbf{A}(t),\mathbf{B}(t))$ is totally controllable on $\mathbb{T}$. An analogous statement exists for SO. \label{def: LTV_structural_controllability}.
\end{defn}

The concepts of SC/SO only state the \textit{existence} of controllable/observable systems of a given structure.  Ruling out the existence of uncontrollable/unobservable systems requires stronger statements, which are supplied by the concepts of \textit{strong} structural controllability/observability (SSC/SSO).  We use the definition of SSC given in \cite{Mayeda_1977}:

\begin{defn}
The system ($\mathbf{A}_0$,$b_0$) is SSC if any system (${\mathbf{A}}_1$,${b}_1$) which has the same structure as ($\mathbf{A}_0$,$b_0$) is completely controllable as long as every free parameter of the matrix $[{\mathbf{A}_1},{b}_1]$ is nonzero.\label{def: strong_struct_cont}
\end{defn}

This is obviously generalizable to a multi-input case by replacing $b,\bar{b}\in\mathbb{R}^{n\times 1}$ with $\mathbf{B},\bar{\mathbf{B}}\in\mathbb{R}^{n\times m}$ and applying the same restriction.  An analagous statement may be made for SSO.  Systems with these properties will be controllable/observable in the sense of Section \ref{subsec: classial_cont_obs} regardless of any errors in their free parameter estimations, so long as these parameters are nonzero.  A similar extension exists for time-varying systems:

\begin{defn}
\eqref{eq: Sample_LTV_sys} of structure $(\mathcal{A},\mathcal{B})$ is strongly structurally controllable if \eqref{eq: Sample_LTV_sys} is totally controllable on $\mathbb{T}$ for every $\mathbf{A}(t)\in\mathcal{A}$ and $\mathbf{B}(t)\in\mathcal{B}$ \cite{Hartung_2013_Sufficient}. \label{def: strong_struct_cont_LTV}
\end{defn}
\section{PBF Thermal Model}\label{sec:thermal_model}
This section defines how basic models of the PBF process can be represented by the system models in \eqref{eq:nonlinear_generic} -- \eqref{eq: Sample_LTI_sys}, and thus a graph $G(\mathbf{A})$, for the different actuation and measurement modes (Sections \ref{sec:intro_actuation} and \ref{sec:intro_measurement}).  

\subsection{PBF model assumptions and construction}
The PBF model used to assess controllability and observability leverages the basic physics, actuator and measurement structure from Section \ref{sec:In}.  We apply several simplifying assumptions that are common in PBF modeling:   

\begin{enumerate}
\setlength{\itemindent}{0em}
\item[\textbf{A1.}] The Biot number for PBF is approximately $Bi=0.01$, thus $q_{\textnormal{conv.}}$ and $q_{\textnormal{rad.}}$ are assumed to be zero \cite{paul_2014}.
\item[\textbf{A2.}] Conduction into the unfused powder is negligible as the loosely packed powder is a poor conductor \cite{rombouts_2005}.  This sets up the insulated Neumann boundary condition at surfaces $\Gamma$, $\nabla T\cdot \hat{\mathbf{n}}=0\textnormal{ }\forall \textnormal{ }\mathbf{v}\in\Gamma$, where $\nabla=\left(\frac{\partial}{\partial x},\frac{\partial}{\partial y},\frac{\partial}{\partial z}\right)$, $\cdot$ is the vector dot product and $\hat{\mathbf{n}}$ is the direction normal to the domain $\Gamma$, as shown in Fig. \ref{fig: PBF_BCs}a.
\item[\textbf{A3.}]  Surfaces $\Lambda$ have constant temperature $T_0$, setting up the Dirichlet boundary condition $T=T_0 \textnormal{ }\forall\textnormal{ } \mathbf{v}\in\Lambda$ and $t$, as shown in Fig. \ref{fig: PBF_BCs}(a). This represents the assumption that the machine baseplate is an ideal heat sink.
\item[\textbf{A4.}] The last, and perhaps most tenuous, assumption is that the top layer is composed of fully-fused metal with a thermal conductivity equal to the bulk conductivity.  This represents the assumption that new material added to the build in the time frames we consider is negligible in comparison to the volume of $V$.  We invoke this assumption to understand the controllability and observability of systems with fixed domains prior to examining the effect of material addition on the problem.  

\end{enumerate}

\subsection{Reduction of PBF dynamics through FEM}

Assumptions \textbf{A1} -- \textbf{A4} reduce \eqref{eq: FluxBalance} to the well-known conductive heat transfer boundary value problem as defined by Fourier's Law:

\begin{equation}
\begin{split}
&\frac{\partial T}{\partial t}=\frac{K}{c\rho}\nabla^2T\textnormal{ }\forall\textnormal{ }\mathbf{v}\in V\\
&T=T_0\textnormal{ }\forall\textnormal{ }\mathbf{v}\in\Lambda\\
&\nabla T\cdot\mathbf{\hat{n}}=0\textnormal{ }\forall\textnormal{ }\mathbf{v}\in\Gamma\\
&\nabla T\cdot\mathbf{\hat{n}}=u(\mathbf{\bar{v}},t)\textnormal{ }\forall\textnormal{ }\mathbf{\bar{v}}\in\Omega.
\end{split}\label{eq: FouriersLaw}
\end{equation}

\noindent $K$ is the material thermal conductivity.  \eqref{eq: FouriersLaw} has no general closed-form solution due to the arbitrarily complex problem domain and boundary conditions imposed by the PBF process.  Several approaches are available to arrive at numeric solutions to \eqref{eq: FouriersLaw}.  We choose FEM to approximate \eqref{eq: FouriersLaw}.  The FEM algorithm is based on energy method solutions to boundary value problems, and thus the FEM solution to \eqref{eq: FouriersLaw} leverages its' \emph{functional}, or \emph{weak} form:

\begin{equation}
\begin{aligned}
\Pi =& \int_V \left( \frac{1}{2}(\nabla T)^\prime \mathbf{\kappa}\nabla T  + \rho c\dot{T}T \right)dV\\
&-\int_{S}\left(u_B T\right) dS,
\end{aligned}
\label{eq:fourier_weak}
\end{equation}

\noindent Where $\mathbf{\kappa}\in\mathbb{R}^{3\times 3}$ is an array that contains directional heat conductivity properties (equal to $K\mathbf{I}_3$ if isotropic), $S$ is the boundary of $V$, and $u_B$ specifies the heat flux on the boundary \cite{Cook89}; heat transfer at the boundary is from both $u(\bar{\mathbf{v}},t)$ at $\Omega$ and the isothermal boundary condition at $\Lambda$.  To help with reader intuition, the first integral captures intra-volume heat transfer and storage at each instant in time (thermal energy captured within the system) and the second integral captures inter-volume heat transfer at each instant in time (thermal energy added to the system).

The FEM is applied to \eqref{eq:fourier_weak} by discretizing $V$ into a set of nodes and elements, as shown in Fig. \ref{fig: PBF_BCs}b. According to Assumption \textbf{A.4}, the nodes are assumed to hold fixed positions at all times, which reduces \eqref{eq:fourier_weak} to the system of ordinary differential equations shown in \eqref{eq: FEM_ODEs}.  $\mathbf{x}$ in \eqref{eq: FEM_ODEs} contains the temperature states at all FEM nodes not on $\Lambda$, since those on $\Lambda$ are constrained to always assume the value $T_0$.  The full derivation of \eqref{eq: FEM_ODEs} is given in Appendix \ref{Appendix: FEM_derivation}.  In \eqref{eq: FEM_ODEs},  the \emph{conductivity matrix} $\mathbf{K}$ captures the conductivity between adjacent nodes, the \emph{capacitance matrix} $\mathbf{M}$ captures the thermal capacitance of each element, and the \emph{load vector} function $\mathbf{R}(t)$ imposes the boundary conditions at the nodes on the boundaries of $V$.  $\mathbf{M}$ and $\mathbf{K}$ are guaranteed to be symmetric and PD \cite{Cook89}. $\mathbf{M}>0$ ensures that $\mathbf{M}^{-1}$ always exists.  To ease the computational burden of inverting $\mathbf{M}$, we elect to use a \emph{lumped mass approximation} in which $\mathbf{M}>0$ is diagonal \cite{ANSYS_2018}.
\begin{equation}
\dot{\mathbf{x}}=-\mathbf{M}^{-1}\mathbf{K}\mathbf{x}+\mathbf{M}^{-1}\mathbf{R}(t).\label{eq: FEM_ODEs}
\end{equation}

In the context of \eqref{eq:nonlinear_generic}, the system dynamics are thus:

\begin{equation}
\begin{aligned}
\dot{\mathbf{x}}&=\mathbf{A}\mathbf{x}+\mathbf{r}(t,u)\\
\mathbf{y} &= \mathbf{g}(t,\mathbf{x}),
\end{aligned}
\label{eq:nonlinear_specific}
\end{equation}

\noindent Where $\mathbf{A}=-\mathbf{M}^{-1}\mathbf{K}$, $\mathbf{r}(t,u)=\mathbf{M}^{-1}\mathbf{R}(t)$, and $\mathbf{g}(t,\mathbf{x})$ is a vector that selects the nodes on $\Omega$ that are visible to measurement, depending on measurement mode (\textbf{M.1} -- \textbf{M.3}).  We now state some properties of $\mathbf{A}$:

\begin{rem}
$\mathbf{M}$ and $\mathbf{K}$ are symmetric and PD; therefore, by \cite{johnson_1977}, $\mathbf{A}$ is Hurwitz. \label{rem: PD_Hurwitz}
\end{rem}

\begin{thm}
\eqref{eq:nonlinear_specific} is asymptotically stable, detectable, and stabilizable.
\end{thm}

\emph{Proof:} LTI system stability, detectability, and stabilizability follow directly from the statement in Remark \ref{rem: PD_Hurwitz} that $\mathbf{A}=-\mathbf{M}^{-1}\mathbf{K}$ is Hurwitz. $\blacksquare$

\begin{thm}
$\mathbf{A}$ has real, negative eigenvalues and is always diagonalizable.  Furthermore, all $[\mathbf{A}]_{ii}\neq0$ and $[\mathbf{A}]_{ij}\neq0\leftrightarrow\mathbf{A}_{ji}\neq0$.\label{thm: eigenstructure_of_A}
\end{thm}

\emph{Proof}: The statements that the eigenvalues of $\mathbf{A}$ are real and negative and that $\mathbf{A}$ is diagonalizable are a direct consequence of the facts that $\mathbf{M}=\text{diag}([\mathbf{M}]_{11},\dots,[\mathbf{M}]_{nn})>0$, $\mathbf{K}=\mathbf{K}^{\prime}>0$ and $\mathbf{A}=-\mathbf{M}^{-1}\mathbf{K}$ (Corollary 7.6.2 of \cite{Horn_2012}).  Since $\mathbf{M}>0$ and $\mathbf{K}>0$, all $[\mathbf{M}]_{ii}>0$ and therefore $[\mathbf{M}^{-1}]_{ii}>0$, and all $[\mathbf{K}]_{ii}>0$ \cite{johnson_1970}.  Since $[\mathbf{A}]_{ij}=-[\mathbf{M}^{-1}]_{ii}[\mathbf{K}]_{ij}$ and $[\mathbf{K}]_{ij}\neq0\leftrightarrow[\mathbf{K}_{ji}]\neq0$, all $[\mathbf{A}]_{ii}\neq0$ and $[\mathbf{A}]_{ij}\neq0\leftrightarrow[\mathbf{A}]_{ji}\neq0$. $\blacksquare$  

\begin{thm}
$\mathbf{A}$ as defined in \eqref{eq: FEM_ODEs} assumes a block diagonal structure if and only if the build geometry $V$ of the system comprises a set of $l$ disconnected structures, $V\coloneqq\left\{V_1,V_2,\dots,V_l\right\}$.
\end{thm}
\emph{Proof}: It is known that ``in the row of $\mathbf{K}$ corresponding to any node $i$ [in the FEM mesh], the nonzero blocks [entries] of $\mathbf{K}$ appear in those columns which locate nodes in the same edge, face, or element as $i$'' \cite{Feng_1983}.  Qualitatively this means that any element $\mathbf{K}_{ij}$ of $\mathbf{K}$ is nonzero when nodes $i$ and $j$ are directly connected within the FEM network.  By the proof of Theorem \ref{thm: eigenstructure_of_A}, $[\mathbf{K}]_{ij}\neq0\leftrightarrow[\mathbf{A}]_{ij}\neq0$, therefore the nonzero entries of $\mathbf{A}$ may be used to identify nodes in direct contact.  Without loss of generality assume that \eqref{eq:nonlinear_specific} is homogeneous ($\mathbf{r}(t,u)=\mathbf{0}$), so that $\dot{\mathbf{x}}=\mathbf{Ax}$.  We use these observations to prove the claim:\par    

\emph{proof (sufficiency)}: Assume that $\mathbf{A}\in\mathbb{R}^{n\times n}$ has a block diagonal structure of $l\leq n$ blocks:

\begin{equation}
    \begin{bmatrix}\dot{\mathbf{x}}_1 \\\dot{\mathbf{x}}_2\\\vdots\\\dot{\mathbf{x}}_l\end{bmatrix}=\begin{bmatrix}\mathbf{A}_1&{\mathbf{0}}&{\cdots}&{\mathbf{0}}\\{\mathbf{0}}&\mathbf{A}_2&{\cdots}&{\mathbf{0}}\\\vdots&\vdots&\ddots&\vdots\\{\mathbf{0}}&{\mathbf{0}}&{\cdots}&\mathbf{A}_l\end{bmatrix}\begin{bmatrix}\mathbf{x}_1\\\mathbf{x}_2\\\vdots\\\mathbf{x}_l\end{bmatrix}.\label{eq: block_diagonal_A}
\end{equation}

\noindent \eqref{eq: block_diagonal_A} shows that entries $[\mathbf{A}]_{ij}$ corresponding to nodes $i$ and $j$ of dissimilar blocks $\mathbf{x}_m$, $\mathbf{x}_r$, $m\neq r$ always take the value $[\mathbf{A}]_{ij}=0$.  By the above observation, this implies that each block $\mathbf{x}_s$, $s=1,\dots,l$ is physically disconnected from all the others.  Therefore each block $\mathbf{x}_s$ represents a physically disconnected structure within the build geometry and hence the build geometry $V$ may be partioned as $V\coloneqq\left\{V_1,\dots,V_l\right\}$.\par

\emph{proof (necessity)}:  Assume that $V$ may be partioned as $V\coloneqq\left\{V_1,\dots,V_l\right\}$.  When constructing \eqref{eq:nonlinear_specific} via FEM, the mesh of each $\mathbf{V}_s$, denoted as $\mathbf{x}_s$, is physically disconnected from the others.  It follows immediately from the above observation that all entries $[\mathbf{A}]_{ij}$ of $\mathbf{A}$ take the value $[\mathbf{A}]_{ij}=0$ whenever nodes $i$ and $j$ belong to $\mathbf{x}_m$, $\mathbf{x}_r$, respectively, $m\neq r$.  $\mathbf{A}$ assumes the structure of \eqref{eq: block_diagonal_A} and is therefore block diagonal $\blacksquare$.





\subsection{Expression of \eqref{eq:nonlinear_specific} for each actuation mode}

$\mathbf{r}(t,u)$ has a different form for each actuation mode applied (Section \ref{sec:intro_actuation}), thus influencing the controllability of the process.  Each of the following subsections derives the form of $\mathbf{r}(t,u)$, leveraging the assumed form of the input function $u(\bar{\mathbf{v}},t)$.  
\subsubsection{Actuation mode \textbf{A.A}}

Actuation mode \textbf{A.A} assumes that the form of $u(\bar{\mathbf{v}},t)$ is arbitrary.  This input structure allows for a convenient simplification of $\mathbf{r}(t,u)$:

\begin{equation}
    \mathbf{r}(t,u)=\mathbf{M}^{-1}\mathbf{R}=\mathbf{M}^{-1}\sum_e\int_{S_e}u(\mathbf{v},t)\mathbf{N}_e^\prime dS_e.\label{eq: Input_A.A_early}
\end{equation}

\noindent Where $\mathbf{N}_e$ are functions which interpolate the value of $T(\mathbf{v},t)$ from the temperatures at the nodes that bound the $e^{th}$ element.  $u(\mathbf{v},t)=u(\bar{\mathbf{v}},t)$ if $\mathbf{v}\in\Omega$ and 0 otherwise.  Details have been relegated to Appendix B.  After simplification is complete, the PBF input takes the form given in \eqref{eq: Simplified_A.A_input}, where the specific meanings of $\mathbf{B}$ and $\mathbf{u}(t)$ are given in Appendix \ref{Appendix: Bu_derivation}.

\begin{equation}
    \mathbf{r}(t,u)\approx\mathbf{B}\mathbf{u}(t).\label{eq: Simplified_A.A_input}
\end{equation}

\subsubsection{Actuation mode \textbf{A.B}}
Under actuation modes \textbf{A.B} and \textbf{A.C}, $u(\mathbf{\bar{v}},t)$ assumes the Gaussian form specified in \eqref{eq: HeatSource}.  \eqref{eq: Input_A.A_early} may therefore be expanded into the form shown in \eqref{eq: Input_A.B,C}.  Here, $\Delta x_i(t)=x_{c,i}(t)-x_{\mathbf{\bar{v}}}$ and $\Delta y_i(t)=y_{c,i}(t)-y_{\mathbf{\bar{v}}}$ for $\mathbf{\bar{v}}=\mathbf{v}\in\Omega$ and time-varying laser centerpoint coordinates $x_{c,i}(t)$ and $y_{c,i}(t)$. 

\begin{equation}
\mathbf{r}(t,\mathbf{\bar{v}},P,\sigma^2)=\mathbf{M}^{-1}\sum_e\int_{S_e}\underbrace{\sum_{i=1}^M\frac{P_i}{\sqrt{2\pi\sigma_i^2}}e^{-\frac{\Delta x_i(t)^2+\Delta y_i(t)^2}{2\sigma_i^2}}\mathbf{N}_e^\prime}_{\mathbf{g}(t,\bar{\mathbf{v}},P,\sigma^2)}dS_e. \label{eq: Input_A.B,C}
\end{equation}

Actuation mode \textbf{A.B} assumes that $P_i,$ $\sigma_i,$ $x_{c,i},$ and $y_{c,i}$ are freely controllable at each instant of time $t$ for all $M$ lasers in the system.  In principle one could define an operating point $\left(P_{0,i},\sigma_{0,i},x_{c0,i},y_{c0,i}\right)$ for all lasers in the system and linearize \eqref{eq: Input_A.B,C}.  However, defining an operating point $\left(x_{c0,i},y_{c0,i}\right)$ is inappropriate given that the laser or e-beam sweeping across the build chamber will always result in large deviations in $x_{c,i}(t)$ and $y_{c,i}(t)$ away from any possible operating point selection.  As such, Actuation mode \textbf{A.B} cannot be accurately expressed in a linear systems framework and must remain expressed as the nonlinear model \eqref{eq:nonlinear_specific}.

\subsubsection{Actuation mode \textbf{A.C}}
As in Actuation mode \textbf{A.B}, in Actuation mode \textbf{A.C} $\mathbf{r}(t,u)$ assumes the form given in \eqref{eq: Input_A.B,C}.  However, under Actuation mode \textbf{A.C} only $P_i\geq0$ and $\sigma_i^2>0$ are available for control.  The laser centroids $(x_{c,i}(t),y_{c,i}(t))$ are treated as model parameters instead of control inputs.  Under these conditions, the integrand $\mathbf{g}(t,\bar{\mathbf{v}},P,\sigma^2)$ of \eqref{eq: Input_A.B,C} is continuously differentiable with respect to $P_i\geq0$ and $\sigma_i^2>0$ for all $M$ active lasers in the system.  Additionally, defining an operating point for these inputs is reasonable.  Leibniz's Integration Rule may therefore be exploited to linearize the integral-valued $\mathbf{r}(t,P,\sigma^2)$ about some operating point $\mathbf{u}_0=\begin{bmatrix}P_{0,1}&\sigma_{0,1}^2&\dots&P_{0,M}&\sigma_{0,M}^2\end{bmatrix}^\prime$.  Doing so yields  $\mathbf{B}(t)\mathbf{\delta u}(t)$ as a first-order approximation of $\mathbf{r}(t,P,\sigma^2)$, where\\ $\mathbf{\delta u}(t)=\begin{bmatrix}\delta P_{1}(t)&\delta\sigma_{1}^2(t)&\dots&\delta P_{M}(t)&\delta\sigma_{M}^2(t)\end{bmatrix}^\prime$ and $\mathbf{B}(t)$ is defined below:

\begin{equation}
\begin{split}
&\mathbf{B}(t)=\mathbf{M}^{-1}\begin{bmatrix}
\frac{\partial\mathbf{r}}{\partial \mathbf{P}}&\frac{\partial\mathbf{r}}{\partial\mathbf{\sigma}^2}
\end{bmatrix}\vline_{\mathbf{u_0}}\\
&\textnormal{ }\begin{bmatrix}
\frac{\partial\mathbf{r}}{\partial \mathbf{P}}&\frac{\partial\mathbf{r}}{\partial\mathbf{\sigma}^2}
\end{bmatrix}\vline_{\mathbf{u_0}}=\\
&\begin{bmatrix}
\left(\sum_e\int_{S_e}\frac{1}{\sqrt{2\pi\sigma_{0,1}^2}}\text{exp}\left(-\frac{\Delta x_1(t)^2+\Delta y_1(t)^2}{2\sigma_{0,1}^2}\right)\mathbf{N}_edS\right)^{\prime}\\
\vdots\\
\left(\sum_e\int_{S_e}\frac{1}{\sqrt{2\pi\sigma_{0,M}^2}}\text{exp}\left(-\frac{\Delta x_M(t)^2+\Delta y_M(t)^2}{2\sigma_{0,M}^2}\right)\mathbf{N}_edS\right)^{\prime}\\
\\
\left(\sum_e\int_{S_e}\frac{P_{0,1}\text{exp}\left(-\frac{\Delta x_1(t)^2+\Delta y_1(t)^2}{2\sigma_{0,1}^2}\right)}{\sqrt{2\pi\sigma_{0,1}^2}}\left(-\frac{\pi}{2\pi\sigma_{0,1}^2}+\frac{\Delta x_1(t)^2+\Delta y_1(t)^2}{2(\sigma_{0,1}^2)^2}\right)\mathbf{N}_edS\right)^{\prime}\\
\vdots\\
\left(\sum_e\int_{S_e}\frac{P_{0,M}\text{exp}\left(-\frac{\Delta x_M(t)^2+\Delta y_M(t)^2}{2\sigma_{0,M}^2}\right)}{\sqrt{2\pi\sigma_{0,M}^2}}\left(-\frac{\pi}{2\pi\sigma_{0,M}^2}+\frac{\Delta x_M(t)^2+\Delta y_M(t)^2}{2(\sigma_{0,M}^2)^2}\right)\mathbf{N}_edS\right)^{\prime}
\end{bmatrix}^\prime\\
&\mathbf{u}(t)=\mathbf{u}_0+\mathbf{\delta u}(t).
\end{split}\label{eq: LinearizedHammersteinInput}
\end{equation}

Denote the elements of $\mathbf{B}(t)$ that correspond to the $i^{th}$ node laying on $\Omega$ as $[\mathbf{B}]_{ij}$.  Define the positive constants $\epsilon_{B,ij}\approx0$ to be the values $[\mathbf{B}]_{ij}$ take whenever node $i$ is not within the laser beam spot.   These matrix entries ``ramp-up'' and ``ramp-down'' to/from their full values given in \eqref{eq: LinearizedHammersteinInput} according to smoothed top-hat functions \cite{Boyd_2005} with arbitrary (but nonzero) ramp up/down times.  This construction acknowledges that the laser beam diameter is described stochastically by $\sigma^2$ and therefore some trivial amount of energy is always spread out among the entire node surface, even if that amount is asymptotically small.

\subsection{Expression of \eqref{eq:nonlinear_specific} for each measurement mode}
$\mathbf{g}(t,\mathbf{x})$ has a different form for each measurement mode,  thus influencing the observability of the process.  Each of the following subsections derives the form of $\mathbf{g}(t,\mathbf{x})$, leveraging the assumed form of the windowing function $h$.

\subsubsection{Measurement mode \textbf{M.1}}
Under Measurement mode \textbf{M.1}, the only available measurement is a spot measurement of ambient temperature through the air or the base plate.  Under Assumptions \textbf{A.1} and \textbf{A.3}, there is negligible heat transfer from the build to the ambient air and there is negligible temperature change in the base plate, respectively.  The measurement $y(t)$ takes the constant value $y(t)=T_0$ (base plate measurement) or $y(t)=T_{amb}$ (ambient air measurement) $\forall$ $t\geq0$, regardless of any temperature dynamics occuring throughout $V$ as described in $\mathbf{x}(t)$.  Therefore no meaningful information regarding $\mathbf{x}(t)$ is captured by these measurements.  Accordingly, under Measurement mode \textbf{M.1}, $y=\mathbf{g}(t,\mathbf{x})$ may be expressed as:

\begin{equation}
    y=\mathbf{0}\mathbf{x}.\label{eq: Output_M.1}
\end{equation}

\begin{rem}
It is trivial to show using the criteria of Section \ref{subsec: classial_cont_obs} that the output relation \eqref{eq: Output_M.1} will produce an unobservable system for any $\mathbf{A}$.\label{remark: unobservable_output_M1}
\end{rem}
 
\subsubsection{Measurement mode \textbf{M.2}}
Under Measurement mode \textbf{M.2} the available measurement is a camera having a fixed FOV.  Therefore, for an assumed-constant FEM mesh, the output of the system $\mathbf{y}=\mathbf{g}(t,\mathbf{x})\in\mathbb{R}^p$ is defined as:

\begin{equation}
\mathbf{y}\equiv\mathbf{Cx}, \label{eq: Output_M.2}
\end{equation}

\noindent Where $\mathbf{C}\in\mathbb{R}^{p\times n}$ selects all nodes on $\Omega$ that lay in the fixed camera FOV to the limit of the camera resolution.  Element $[\mathbf{C}]_{ij}$ is only nonzero if output $\mathbf{y}_i$ corresponds to state $\mathbf{x}_j$, and then assumes the value 1.
In the case where the measurement is a spot measurement from a pyrometer centered at $\left(\bar{x},\bar{y}, z=\bar{z}\right)$, $\mathbf{C}$ selects the single node on $\Omega$ that is closest to $\left(\bar{x},\bar{y},z=\bar{z}\right)$.

\subsubsection{Measurement mode \textbf{M.3}}
Under Measurement mode \textbf{M.3}, the available measurement is a camera mounted coaxially with the laser that has a FOV moving with the laser centerpoint.  The laser centerpoint is treated as a governing parameter of the system output.  Therefore, for an assumed-constant FEM mesh, the output of the system $\mathbf{y}=\mathbf{g}(t,\mathbf{x})\in\mathbb{R}^p$ is defined as:

\begin{equation}
    \mathbf{y}\equiv\mathbf{C}(t)\mathbf{x}.\label{eq: SystemOutput_LTV}
\end{equation}

\noindent Here, $\mathbf{C}(t)\in\mathbb{R}^{p\times n}$ is a time-varying selection matrix that selects the set of nodes on $\Omega$ being observed by the moving melt pool camera FOV to the limit of the camera resolution.  The theory of Section \ref{sec: cont_obs_cases_2_4} requires that $\mathbf{C}(t)$ be smooth.  We represent the process of a given node on $\Omega$ entering, exiting, and laying within the camera FOV by populating the nonzero entries of $\mathbf{C}(t)$ with smoothed ``top-hat'' functions of the type discussed in \cite{Boyd_2005}.  Whenever node $j$ is not within the melt pool camera FOV, all corresponding entries $[\mathbf{C}]_{ij}$ of $\mathbf{C}$ assume small positive constant values $\epsilon_{C,ij}\approx0$.  This construction acknowledges that the camera always receives some trivial quantity of light that is sourced from any location on $\Omega$ due to reflections and diffuse emission of radiation, even if that amount is asymptotically small.  As in Measurement mode \textbf{M.2}, in the event that the available measurement is a coaxially-mounted pyrometer focused on the point $\left(x_{c,i}+\bar{x},y_{c,i}+\bar{y}\right)$, $\mathbf{C}(t)$ selects the single node on $\Omega$ that is closest to $\left(x_{c,i}+\bar{x},y_{c,i}+\bar{y}\right)$.

\subsection{Linear system construction}
The $\mathbf{B}(t)\mathbf{u}$ and $\mathbf{C}(t)\mathbf{x}$ input-output relationships derived in the previous sections are combined with \eqref{eq:nonlinear_specific} to produce four linear systems:
\begin{enumerate}
    \item Case 1: Actuation mode \textbf{A.A} and Measurement mode \textbf{M.2}
          \begin{equation}
          \begin{split}
           &\mathbf{\dot{x}}=\mathbf{Ax}+\mathbf{Bu}\\
           &\mathbf{y=Cx}.
           \end{split}\label{eq: LTI_System}
           \end{equation}
    \item Case 2: Actuation mode \textbf{A.A} and Measurement mode \textbf{M.3}
           \begin{equation}
          \begin{split}
           &\mathbf{\dot{x}}=\mathbf{Ax}+\mathbf{Bu}\\
           &\mathbf{y=C}(t)\mathbf{x}.
           \end{split}\label{eq: LTV_System_Case2}
           \end{equation}
    \item Case 3: Actuation mode \textbf{A.C} and Measurement mode \textbf{M.2}
        \begin{equation}
          \begin{split}
           &\mathbf{\dot{x}}=\mathbf{Ax}+\mathbf{B}(t)\mathbf{u}\\
           &\mathbf{y=Cx}.
           \end{split}\label{eq: LTV_System_Case3}
           \end{equation}
    \item Case 4: Actuation mode \textbf{A.C} and Measurement mode \textbf{M.3}
            \begin{equation}
          \begin{split}
           &\mathbf{\dot{x}}=\mathbf{Ax}+\mathbf{B}(t)\mathbf{u}\\
           &\mathbf{y=C}(t)\mathbf{x}.
           \end{split}\label{eq: LTV_System_Case4}
           \end{equation}
\end{enumerate}
These four cases represent the two Actuation modes that may be reasonably linearized and the two Measurement modes that contain useful information about the system dynamics.  Measurement mode \textbf{M.1} was discarded due to always producing an unobservable system by Remark \ref{remark: unobservable_output_M1}.  In Cases 3 and 4, the system input is $\mathbf{u}=[\delta P_1,\delta \sigma^2_1,\dots,\delta P_M,\delta\sigma^2_M]$, the deviations in nominal laser power/variance away from their nominal values. 
Case 1 is the only LTI system among the four, with Cases 2-4 being time-variant in the $\mathbf{B}(t)$ and/or $\mathbf{C}(t)$ matrices.

\begin{figure*} [!tb]
\centering
\includegraphics[width=0.90\textwidth]{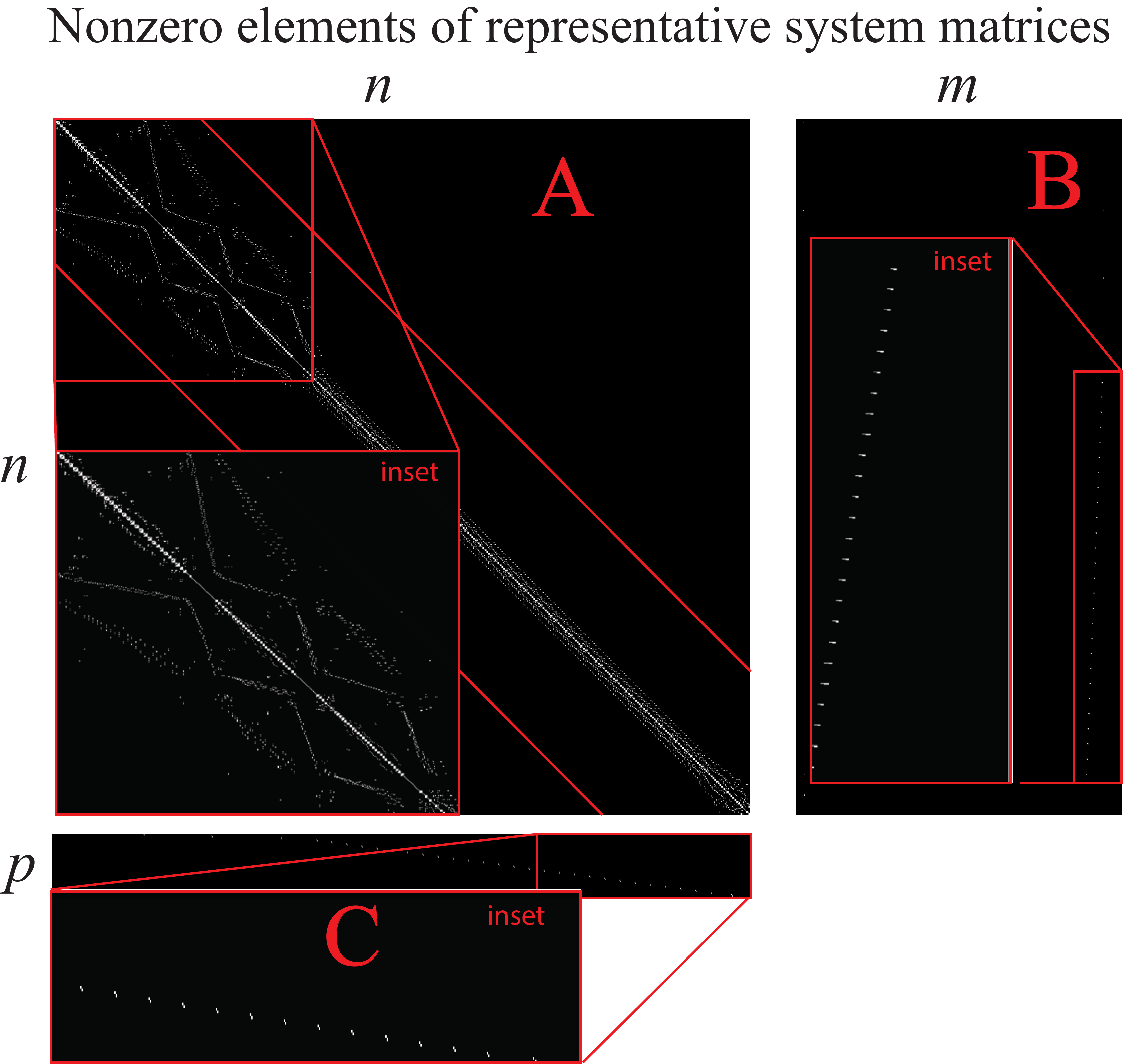}
\caption{Nonzero element locations of representative $\mathbf{A} \in \mathbb{R}^{n \times n} $, $\mathbf{B}\in\mathbb{R}^{n\times m}$, and $\mathbf{C} \in \mathbb{R}^{p \times n}$ matrices.  For this system, $n=1200$, $m=540$, and $p=63$.  The red bars on the $\mathbf{A}$ matrix indicate the approximate bounds of the nonzero element band, which is 239 entries wide.  Insets provide a closer view of the pattern of non-zero elements of the matrices.}
\label{fig: SampleMatrices}
\end{figure*}

Sample matrices for a small representative system of Case 1 are displayed in Fig. \ref{fig: SampleMatrices}, which demonstrates a notable amount of sparseness.  The majority of $\mathbf{A}$ is zero because only a select few nodes are adjacent to any given node in the FEM mesh.  $\mathbf{B}$ models the system input via nodes on $\Omega$ and $\mathbf{C}$ identifies output nodes from those laying on $\Omega$.  For typical part geometries, the majority of the nodes are not on $\Omega$, resulting in $\mathbf{B}$ and $\mathbf{C}$ arrays which are sparse as well.  

\section{Controllability and Observability, Case 1}\label{sec: cont_obs_case_1}

\subsection{Structural and classical controllability/observability}
Our first result is a set of conditions on the controllability/observability of Case 1 for a $V$ without disconnected structures. 

\begin{thm}
Suppose that Case 1 models a build domain $V$ with no disconnected structures and that $V$ contains at least one node on $\Omega$.  Then Case 1 is structurally controllable and structurally observable in the sense of Definition \ref{def: structural_controllability}.  Furthermore, systems of Case 1 are controllable and observable in the sense of Section \ref{subsec: classial_cont_obs} if $\mathbf{A}$ has all distinct eigenvalues.\label{thm: LTI_cont_obs_conditions}
\end{thm}

\emph{Proof}:  We prove the claim explicitly for controllability and then show that the proof for observability proceeds similarly.  We first answer the question of existence -- Is it possible for systems constructed according to the claim to be controllable?  This is answered with the notion of structural controllability.  By Theorem \ref{thm: eigenstructure_of_A} all $[\mathbf{A}]_{ii}\neq0$.  It follows then from \cite{Cowan_2012} that Case 1 is structurally controllable from a single driver node placed on the Power Dominating Set (PDS) of the FEM network.  The PDS of a network is the smallest set of nodes such that all other nodes of the network are ``downstream'' from at least one node in the network.  Specifying that $V$ contains no disconnected structures means that every node in the FEM mesh of $V$ is connected to at least one other node.  By Theorem \ref{thm: eigenstructure_of_A}, every node in $\mathbf{A}$ is connected \textit{bidirectionally} to at least one other node, $[\mathbf{A}]_{ij}\neq0\leftrightarrow[\mathbf{A}]_{ji}\neq0$, and therefore any node $i$ is downstream of any other node $j$ since a path may always be drawn from node $i$ to node $j$.  This implies that the PDS of the network of Case 1 can be defined as any node in the network and therefore Case 1 may be controlled by controlling this single node, hereafter referred to as the system \textit{driver node} and counted with the label $N_D=1$.  $N_D$ generally represents the minimum number of independent driver nodes needed to control the system, which in terms of $\mathbf{B}$ of Case 1 means that $\text{rank}(\mathbf{B})\geq N_D$.  The only nodes eligible for control are those which lay on $\Omega$.  By assumption at least one node on $\Omega$ exists, denoted as node $d$.  By selecting this node for control, $[\mathbf{B}]_{dj}\neq0$, $j=1,\dots,m$, we render Case 1 structurally controllable. Therefore at least one controllable system of the same structure as Case 1 exists.  

Having shown that a system constructed according to Case 1 that is controllable with $N_D=1$ exists, we wish to evaluate under what conditions it appears.  Let $\delta(\lambda_i)$ denote the algebraic multiplicity of the eigenvalue $\lambda_i$ of $\mathbf{A}$, defined as the amount of times $\lambda_i$ is repeated in the set of all eigenvalues of $\mathbf{A}, \text{spec}(\mathbf{A})$.  Let $\mu(\lambda_i)$ denote the geometric multiplicity of $\lambda_i$, defined as the number of linearly independent eigenvectors associated with $\lambda_i$, or equivalently the number of Jordan blocks associated with $\lambda_i$ when $\mathbf{A}$ is placed in Jordan canonical form. \cite{Yuan_2013} shows that the minimum number of driver nodes $N_D$ needed to control a network in the sense of Section \ref{subsec: classial_cont_obs} is given by $N_D=\max_i\left\{\mu(\lambda_i)\right\}$.  We have shown the existence of systems constructed according to the claim ($N_D=1$) that are controllable, therefore there exist $\mathbf{A}$ of Case 1 for which $\max_i\left\{\mu(\lambda_i)\right\}=1$ and these systems are controllable.  This requirement corresponds to two conditions on $\mathbf{A}$:

\begin{enumerate}
    \item All eigenvalues of $\mathbf{A}$ are distinct and therefore $\mu(\lambda)=1$ $\forall$ $\lambda\in\text{spec}(\mathbf{A})$.
    \item Any repeated eigenvalue ($\delta(\lambda_i)>1$) has the potential for $\mu(\lambda_i)>1$.  For $\mu(\lambda_i)=1$ in this case, $\mathbf{A}$ cannot be diagonalizable since its' Jordan canonical form must contain a single ($\delta(\lambda_i)\times\delta(\lambda_i)$) Jordan block that corresponds to $\lambda_i$.
\end{enumerate}

By Theorem \ref{thm: eigenstructure_of_A}, the presence of repeated eigenvalues cannot result in $\text{max}_i\left\{\mu(\lambda_i)\right\}=1$ since $\mathbf{A}$ is always diagonalizable.  Therefore systems constructed according to Case 1 are controllable with $N_D=1$ if they have distinct eigenvalues, and the claim for controllability is proven.

The structural observability of $G(\mathbf{A})$ is equivalent to the structural controllability of $G(\mathbf{A}^\prime )$ \cite{Liu_YY_2013}.  It is trivial to show that every property of $\mathbf{A}$ given in Theorem \ref{thm: eigenstructure_of_A} holds for $\mathbf{A}^{\prime}$; the proof for observability proceeds similarly to that for controllability. $\blacksquare$

We now extend this result to $V$ with a set of disconnected structures. 

\begin{thm}
Let $V$ be partioned into a set of disconnected structures, $V\coloneqq\left\{V_1,V_2,\dots,V_l\right\}$, $l\neq n$.  Then Case 1 is structurally controllable/observable in the sense of Definition \ref{def: structural_controllability} if at least one node exists on the exposed build surface of each $V_i$, $i=1,\dots,l$.  Systems of this structure are controllable and observable in the sense of Section \ref{subsec: classial_cont_obs} if all $\mathbf{A}_s$ have all distinct eigenvalues.\label{thm: LTI_controllability_disc_struct}
\end{thm}

\emph{Proof}: As proved by Theorem 1, Case 1 assumes the following structure if and only if $V$ contains mutually disconnected structures:

\begin{equation}
    \begin{split}
            &\begin{bmatrix}\dot{\mathbf{x}}_1\\\dot{\mathbf{x}}_2\\\vdots\\\dot{\mathbf{x}}_l\end{bmatrix}=\begin{bmatrix}\mathbf{A}_1&{\mathbf{0}}&{\cdots}&{\mathbf{0}}\\{\mathbf{0}}&\mathbf{A}_2&{\cdots}&{\mathbf{0}}\\\vdots&\vdots&\ddots&\vdots\\{\mathbf{0}}&{\mathbf{0}}&{\cdots}&\mathbf{A}_l\end{bmatrix}\begin{bmatrix}\mathbf{x}_1\\\mathbf{x}_2\\\vdots\\\mathbf{x}_l\end{bmatrix}+\begin{bmatrix}\mathbf{B}_1\\\mathbf{B}_2\\\vdots\\\mathbf{B}_l\end{bmatrix}\mathbf{u}\\
            &\mathbf{y}=\mathbf{y}_1+\mathbf{y}_2+\cdots+\mathbf{y}_l=\begin{bmatrix}\mathbf{C}_1&\mathbf{C}_2&\cdots&\mathbf{C}_l\end{bmatrix}\begin{bmatrix}\mathbf{x}_1\\\mathbf{x}_2\\\vdots\\\mathbf{x}_l\end{bmatrix}.
    \end{split}\label{eq: Case_1_block_diagonal}
\end{equation}

The controllability/observability of \eqref{eq: Case_1_block_diagonal} is equivalent to assessing the controllability and observability of each disconnected constituent subsystem individually:

\begin{align}
        {} && {} && \begin{split}
        &\dot{\mathbf{x}}_s=\mathbf{A}_s\mathbf{x}_s+\mathbf{B}_s\mathbf{u}\\
        &\mathbf{y}_s=\mathbf{C}_i\mathbf{x}_s
    \end{split}&&s=1,\dots,l\label{eq: subsystem_i}
\end{align}

\noindent \eqref{eq: subsystem_i} comprises the model structure of Case 1 for each build $V_s$ that contains no mutually disconnected structures, therefore by Theorem \ref{thm: LTI_cont_obs_conditions} each is both structurally controllable/observable if there is at least one node on their exposed surfaces, and controllable/observable in the sense of Section \ref{subsec: classial_cont_obs} if all eigenvalues of each $\mathbf{A}_s$ are distinct.  $\blacksquare$ 

Having placed conditions for which Case 1 is controllable/observable from a single node on $\Omega$, we now place a crucial limitation on this result:

\begin{rem}
If $\mathbf{A}$ is not controllable/observable in the sense of Section \ref{subsec: classial_cont_obs} from a single node then we cannot necessarily restore controllability/observability by simply adding more driver/observer nodes to exposed face $\Omega$.  \cite{Yuan_2013} shows that driver nodes in a network must be placed where the matrix pencil $(\lambda^M\mathbf{I}_n-\mathbf{A})$ loses rank, where $\lambda^M$ is the eigenvalue with maximal $\mu(\lambda_i)=\mu^M$.  There is no guarantee that these $N_D>1$ required driver nodes all lay on $\Omega$ if $\mu^M>1$ and if they do not, then controllability of Case 1 is impossible.
\end{rem}

\subsection{Strong structural controllability/observability}

Having shown a condition for which Case 1 admits a controllable and observable $(\mathbf{A},\mathbf{B},\mathbf{C})$, we now ask if it is possible for this condition to fail.  To answer this question we turn to the concepts of SSC and SSO; as given in Definition \ref{def: strong_struct_cont}. We first approach the problem of SSC using the necessary and sufficient set of conditions supplied in \cite{Reissig_2014}, using the terminology of Section \ref{subsec: struc_cont_obs}:

\begin{thm}
An LTI system described by the graph $G(\mathbf{A,B})$ is SSC if and only if both the following two conditions hold \cite{Reissig_2014}:
\begin{itemize}
    \item [($G_0$)] For every non-empty subset $N$ of non-input nodes in the graph there exists a node $j$ in the graph (including input nodes), such that $N$ contains exactly one successor of $j$.
    \begin{itemize}
        \item Successor: Let there be a directed edge from node $v$ to node $w$ in $G(\mathbf{A,B})$.  Then $w$ is a successor of $v$.
    \end{itemize}
    \item [($G_1$)] For every non-empty subset $N$ of non-input nodes in the graph such that all predecessors of $N$ are contained in $N$, there exists a node $j$ in the graph (including input nodes) that is not in $N$ for which $N$ contains exactly one successor of $j$.
        \begin{itemize}
        \item Predecessor: Let there be a directed edge from node $v$ to node $w$ in $G(\mathbf{A,B})$.  Then $v$ is a predecessor of $w$.
    \end{itemize}
\end{itemize}
\label{thm: SSC_LTI_criterion}
\end{thm}

\begin{figure} [!tb]
\centering
\includegraphics[width=1.0\textwidth]{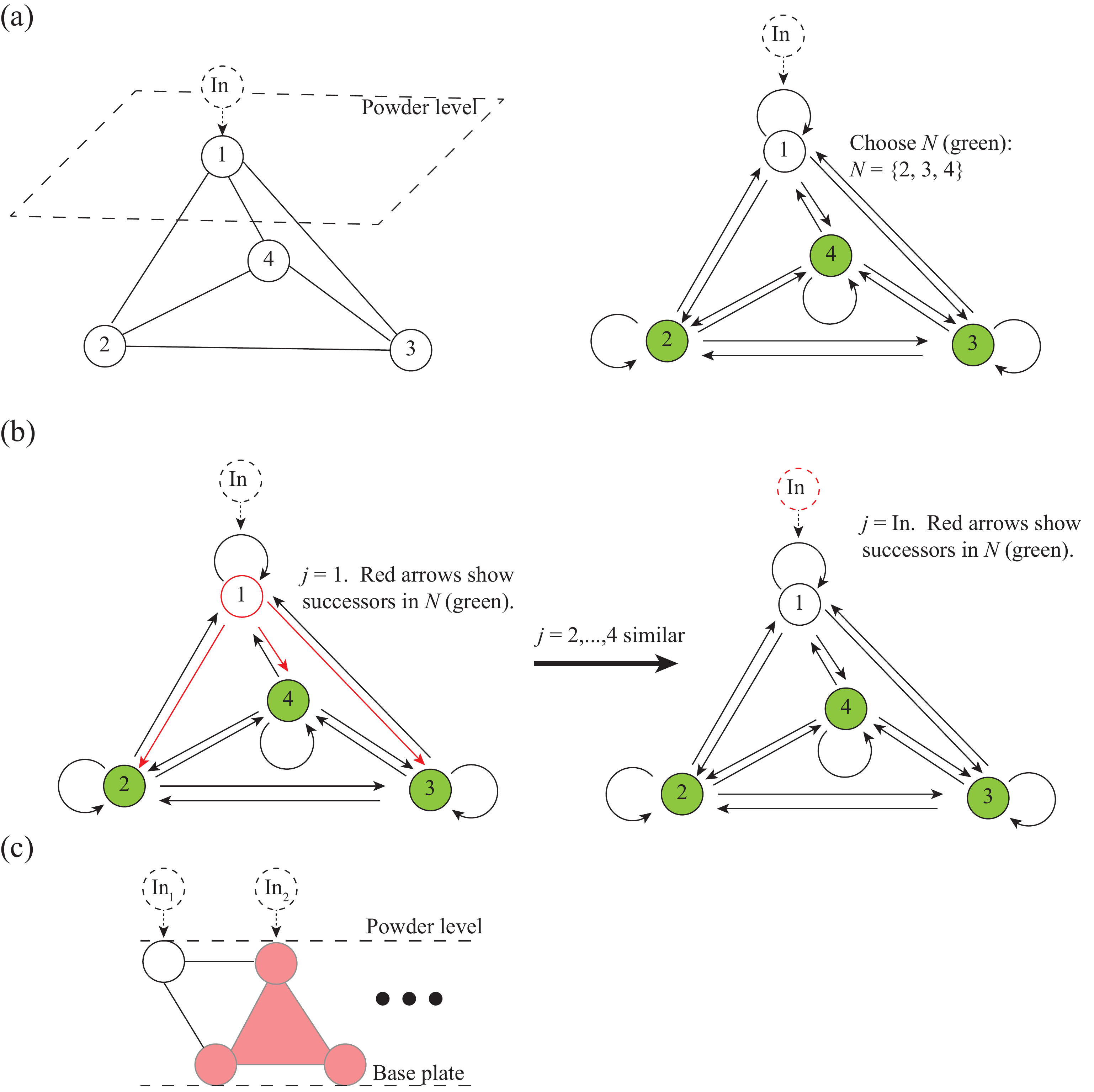}
\caption{\textbf{Failure of criterion $G_0$ of Theorem \ref{thm: SSC_LTI_criterion}}: (a) Single element mesh for 3D build domain $V$ constructed according to Case 1, and sample set $N$ of non-input nodes.  (b) Demonstration that no node $j$ exists such that exactly one successor of $j$ is contained in $N$.  (c) Cross-sectional view of $G(\mathbf{A},\mathbf{B})$ corresponding to a track of material having nonzero contact area with the base plate.  Simplest possible physically realizable build domain.  More complicated domains structured similarly. }
\label{fig: SSC_fig}
\end{figure}

From Theorem \ref{thm: SSC_LTI_criterion} we may draw the following conclusion:

\begin{thm}
It is impossible for Case 1 as constructed from a physically realizable $V$ to be SSC or SSO in the sense of Definition \ref{def: strong_struct_cont}.\label{thm: LTI_SSC_SSO}
\end{thm}
\emph{Proof}:  Fig. \ref{fig: SSC_fig}a shows the graph corresponding to the simplest possible 3D FEM mesh realizable by Case 1.  Since the two criterion of Theorem \ref{thm: SSC_LTI_criterion} must hold for all subsets of $N$ non-input nodes, we choose the $N$ shown in Fig. \ref{fig: SSC_fig}.  We check the validity of Criterion $(G_0)$ as shown in Fig. \ref{fig: SSC_fig}b.  It is clear that varying our choice of $j$ has two outcomes:  There are three successors of $j$ in $N$ if $j$ is chosen among the non-input nodes of the graph, or there are no successors of $j$ in $N$ if $j$ is chosen to be the input node.  Criterion $(G_0)$ fails and the system is not SSC.  

As shown in Fig. \ref{fig: SSC_fig}c, any FEM mesh for build domains that have nonzero contact area with the base plate (physically realizable build domains) will contain elements of this structure and/or elements that are entirely subsurface.  This is because for there to be nonzero surface area in contact with the base plate, at least one element must have an entire face beneath the surface, leaving at most one node in contact with the surface.  Therefore it is guaranteed that Criterion ($G_0$) of Theorem \ref{thm: SSC_LTI_criterion} fails for any $\mathbf{A}$ of Case 1 generated from a physically realizable $V$, since the node set of Fig. \ref{fig: SSC_fig}a must always be present.  Therefore Case 1 cannot be SSC.  The proof for SSO proceeds similarly because the question of controllability of $G(\mathbf{A},\mathbf{C})$ is equivalent to the question of observability of $G(\mathbf{A}^{\prime},\mathbf{C}^{\prime})$ \cite{Liu_YY_2013}. $\blacksquare$

This result shows that it is possible to generate systems of Case 1 that are neither controllable nor observable.  This loss of controllability/observability is not necessarily contingent on a parameter misestimation, it may occur even for systems of certain geometry for which parameters known to infinite precision.
\section{Controllability and Observability, Cases 2-4}\label{sec: cont_obs_cases_2_4}
\subsection{Structural controllability/observability}
We now assess under what conditions Cases 2-4 are controllable and observable in the sense of Section \ref{subsec: classial_cont_obs}.  Throughout this section we leverage the fact that the patterns of $(\mathbf{A}(t),\mathbf{B}(t),\mathbf{C}(t))$ defined in Cases 2-4 are constant.  We begin by verifying the existence of controllable/observable systems of the structures given by Cases 2-4 with the notions of structural controllability/observability:

\begin{thm}
The LTV systems Cases 2-4 are structurally controllable and structurally observable in the sense of Definition \ref{def: LTV_structural_controllability}.\label{thm: LTV_struct_cont_obs}
\end{thm}
\emph{proof}:  Define an LTI system ($\mathbf{A},\mathbf{B}_1,\mathbf{C}_1$) such that $\mathbf{B}_1$ and $\mathbf{C}_1$ have the same patterns as $\mathbf{B}(t)$ and $\mathbf{C}(t)$, respectively, following Definition \ref{def: LTV_structural_controllability}.  Doing so constructs an LTI system according to Case 1.  By Theorems \ref{thm: LTI_cont_obs_conditions} and \ref{thm: LTI_controllability_disc_struct}, there exists at least one set of matrices having the patterns of $\mathbf{B}(t)$ and $\mathbf{C}(t)$ such that the resulting system $(\mathbf{A},\mathbf{B}(t),\mathbf{C}(t))$ is controllable and observable.  Therefore, Cases 2-4 are structurally controllable and structurally observable. $\blacksquare$

\subsection{Strong structural controllability/observability}
We next assess whether uncontrollable/unobservable systems contructed according to Cases 2-4 exist.  \cite{Hartung_2013_Necessary} supplies a necessary condition for SSC of \eqref{eq: Sample_LTV_sys}:

\begin{thm}
A LTV system with matrices $(\mathbf{A}(t),\mathbf{B}(t))$, $\mathbf{A}(t)\in\mathcal{A}$, $\mathbf{B}(t)\in\mathcal{B}$ is SSC within any specified time interval only if the class of linear time-invariant systems $(\mathcal{A}+\mathcal{I}_n,\mathcal{B})$ is also SSC. Here, $\mathcal{I}_n$ represents a diagonal $(n\times n)$ matrix in which all of the diagonal elements are allowed to freely vary \cite{Hartung_2013_Necessary}.\label{thm: SSC_LTV}
\end{thm}

\begin{thm}
It follows directly from Theorem \ref{thm: SSC_LTV} that Cases 2-4 cannot be SSC within any time interval in the sense of Definition 1.\label{thm: Cases_2-4_SSC}
\end{thm}

\emph{proof}: Let $\mathcal{A}_1\coloneqq\mathcal{A}+\mathcal{I}_n$.  Under this definition, $\mathbf{A}_1(t)\in\mathcal{A}_1$ is defined as:

\begin{equation}
    \begin{split}
        &[\mathbf{A}]_{1,ij}=[\mathbf{A}]_{ij},\: i\neq j\\
        &[\mathbf{A}]_{1,ii}=[\mathbf{A}]_{ii}+\delta_{ii}.
    \end{split}\label{eq: A_LTV_cont_def}
\end{equation}

We assume that all nonzero elements $[\mathbf{A}]_{ij}$ (including all $[\mathbf{A}]_{ii}$) are freely variable, and that $\delta_{ij}$ is a freely variable scalar that is independent of $[\mathbf{A}]_{ii}$.  This precisely defines the diagonal elements of $\mathbf{A}_1$ for all LTI systems defined with structures according to Cases 2-4 and belonging to the class $(\mathcal{A}+\mathcal{I}_n,\mathcal{B})$.    

We test for SSC in the sense of Definition 1 for LTI systems belonging to the class $(\mathcal{A}+\mathcal{I}_n,\mathcal{B})$ by assuming that all indeterminate entries of $\mathbf{A}_1$ are nonzero.  $\mathbf{A}_1$ and $\mathbf{A}$ have the same structure, as is made clear from \eqref{eq: A_LTV_cont_def}, therefore we analyze the SSC of LTI systems belonging to the class $(\mathcal{A}_1,\mathcal{B})$ under the framework of Case 1.  We do so by invoking Theorem \ref{thm: LTI_SSC_SSO} and concluding that these systems cannot be SSC $\blacksquare$.

We now show that the problem of SSO is dual to that of SSC by presenting the associating check for SSO of LTV systems:

\begin{thm}
The following two statements are equivalent \cite{Reissig_2014}:
\begin{enumerate}
    \item Every LTV system with $\mathbf{A}\in\mathcal{A}$ and $\mathbf{C}\in\bar{\mathcal{C}}$ for $t_0\leq t\leq t_1$ is observable.
    \item Every LTI system with output pattern $(\mathcal{A}+\mathcal{I}_n,\bar{\mathcal{C}})$ is observable.
\end{enumerate}\label{thm: LTV_SSO}
\end{thm}

\begin{thm}
It follows directly from Theorem \ref{thm: LTV_SSO} that Cases 2-4 cannot be SSO.
\end{thm}
\emph{proof}: The statement of Theorem \ref{thm: LTV_SSO} is obviously dual to that of Theorem \ref{thm: SSC_LTV}.  The proof proceeds similarly to that of Theorem \ref{thm: Cases_2-4_SSC}.

\section{Energy considerations for control and observation}\label{sec: control energy}
The previous sections gave statements on the controllability and observability of Cases 1-4.  The weakness of controllability and observability is that these concepts are binary measurements.  They only assess if driving the system from any state to any other state in finite time is possible, or if reconstructing the system state from input/output measurements in finite time is possible.  No information regarding the practicality of control or observation is captured.  This information is of critical importance to the control engineer because there may exist certain states in a fully controllable/observable system that cannot be reached or reconstructed in practice due to the energy demands placed on the system exceeding the operating ranges of the available actuators and sensors, respectively.  In this section we demonstrate through case studies the relative degree of ``difficulty'' in driving a controllable system to different states in finite time and estimating the state of an observable system in finite time.

Suppose that a discrete-time LTI system is to be driven to some final state $\mathbf{x}_f[k=K]$ with a given input signal $\left\{\mathbf{u}[k]\right\}_{k=0}^{K}$ and generating an output signal $\left\{\mathbf{y}[k]\right\}_{k=0}^K$.  It is well known that the minimum energy needed to reach $\mathbf{x}_f[K]$ among all possible $\left\{\mathbf{u}\right\}_{k=0}^{K}$, $\left\{\mathbf{u}_{min}\right\}_{k=0}^{K}$, is given by \eqref{eq: min_control_energy} \cite{Pasqualetti_2014}:

\begin{equation}
    E_{\text{min},c}=\sum_{k=0}^{K}||\mathbf{u}_{min}[k]||_2^2=\mathbf{x}_f^{\prime}\mathbf{W}_c^{-1}\mathbf{x}_f.\label{eq: min_control_energy}
\end{equation}

In \eqref{eq: min_control_energy}, $\mathbf{W}_c$ represents the discrete-time analogue to the controllability gramian given in Section \ref{sec:Preliminaries}.  Similarly, the energy absorbed by the outputs $\left\{\mathbf{y}[k]\right\}_{k=0}^{K}$ over the course of reconstructing (observing) $\mathbf{x}_f[K]$ is given by \eqref{eq: min_observe_energy} \cite{Pasqualetti_2014}:

\begin{equation}
    E_{obs}=\sum_{k=0}^{K}||\mathbf{y}[k]||_2^2=\mathbf{x}_f^{\prime}\mathbf{W}_o\mathbf{x}_f.\label{eq: min_observe_energy}
\end{equation}

In \eqref{eq: min_observe_energy}, $\mathbf{W}_o$ represents the discrete-time analogue to the observability gramian given in Section \ref{sec:Preliminaries}.  

\subsection{Required controller energy, case study}

We now demonstrate the practicality of control with case studies.  Unfortunately, \eqref{eq: min_control_energy} is not practical to assess $E_{min,c}$ for Cases 1-4 due to the inversion of $\mathbf{W}_c$.  Cases 1-4 typically comprise several hundred to several thousand nodes and $\mathbf{W}_c$ is accordingly ill-conditioned, which produces substantial numerical error during the matrix inversion process.  We instead use the methodology of \cite{zhao_2016}, which we now summarize.  Given an $n$-dimensional discrete-time system $\mathbf{x}[k+1]=\mathbf{A}\mathbf{x}[k]+\mathbf{B}[k]\mathbf{u}[k]$, the desired state $\mathbf{x}_{f}[K]$ may be expressed as

\begin{equation}
    \begin{split}
        &\mathbf{x}_f[K]=\sum_{i=1}^{n}\eta_i\mathbf{V}_i\\
        &\eta_i\coloneqq\sum_{k=0}^{K}\lambda_i^{K-k-1}\text{row}_i\left(\mathbf{V}^{-1}\mathbf{B}[k]\right)\mathbf{u}[k].
    \end{split}\label{eq: controllability_modes}
\end{equation}

\noindent Where $\lambda_i$ are the eigenvalues of $\mathbf{A}$ and $\mathbf{V}=[\mathbf{V}_1,\dots,\mathbf{V}_i,\dots,\mathbf{V}_n]$ is the matrix of associated eigenvectors.  The operator $\text{row}_i(\cdot)$ selects the $i^{th}$ row of a given matrix.  \cite{zhao_2016} assumes that $\mathbf{A}$ is diagonalizable and therefore that $\mathbf{V}$ is invertible.  Each $\eta_i$ of \eqref{eq: controllability_modes} is bounded \cite{zhao_2016}:

\begin{equation}
    \eta_i\leq\eta_i^{*}\coloneqq \sqrt{ \sum_{k=0}^{K}\lambda_i^{2(K-k-1)}||\text{row}_i\left(\mathbf{V}^{-1}\mathbf{B}[k]\right)|| }.\label{eq: eta_bound}
\end{equation}

\eqref{eq: controllability_modes} shows that driving the system to $\mathbf{x}_f[K]$ drives the system along the set of configurations specified by each $\mathbf{V}_i$ simultaneously, with the ``distance'' the system is driven along each $\mathbf{V}_i$ governed by the weight $\eta_i$.  The bound of \eqref{eq: eta_bound} therefore specifies the maximum ``distance'' the system may be driven towards configuration $\mathbf{V}_i$.  By deriving bounds for the weights on this modal decomposition of $\mathbf{x}_f[K]$ we identify the ``preferred'' configurations of the system as those $\mathbf{V}_i$ which have the largest $\eta_i^{*}$.  We reason that less control energy must be expended to reach these ``preferred'' configurations than reaching configurations with small $\eta_i^{*}$.  

\begin{figure*} [!tb]
\centering
\includegraphics[width=0.95\textwidth]{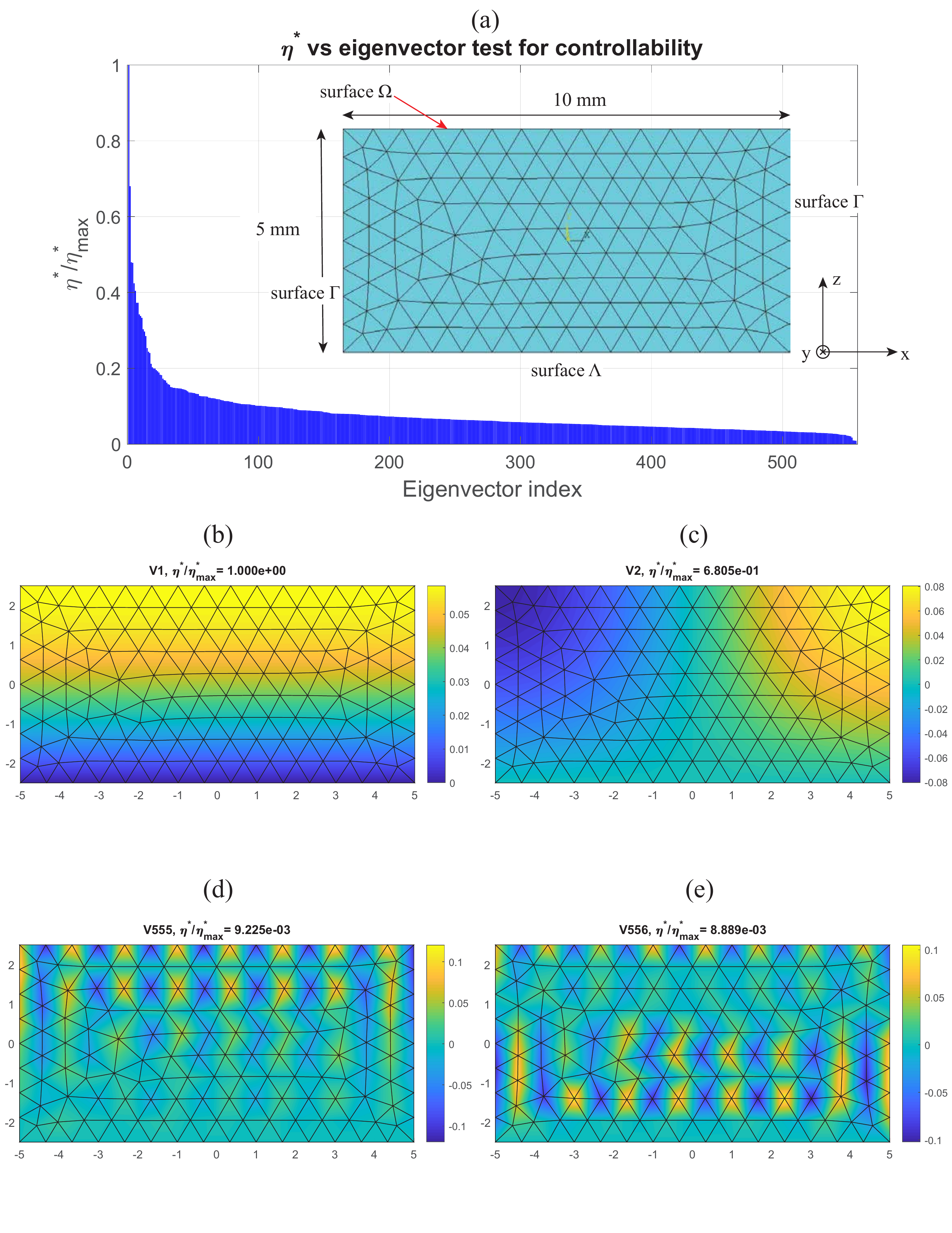}
\caption{\textbf{Controllability modes for simple 2D rectangular mesh}. (a) Geometry of mesh and listing of $\eta_i^{*}$ for all $\mathbf{V}_i$ in decreasing order. Surface $\Omega$ controlled by actuators.  (b) Configuration $\mathbf{V}_1$ associated with $\eta_1^{*}=\eta_{\text{max}}^{*}$. (c) Configuration $\mathbf{V}_2$ associated with $\eta_2^{*}$.  (d) Configuration $\mathbf{V}_{n-1}$ associated with $\eta_{n-1}^{*}$.  (e) Configuration $\mathbf{V}_n$ associated with $\eta_n^{*}=\eta_{\text{min}}^{*}$. }
\label{fig: eta_samples_rectangle}
\end{figure*}

\begin{figure*} [!tb]
\centering
\includegraphics[width=0.93\textwidth]{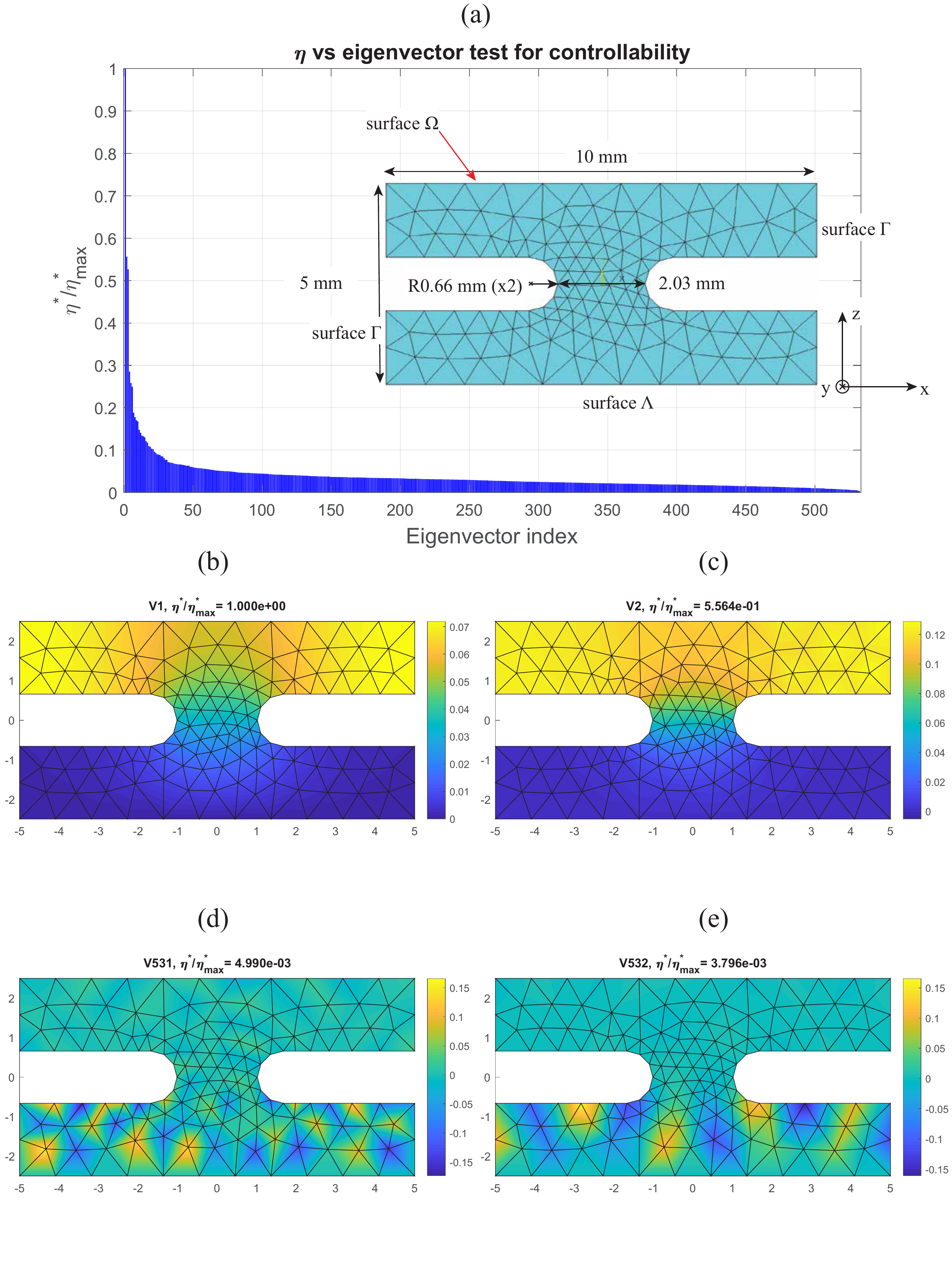}
\caption{\textbf{Controllability modes for simple 2D spool-shaped mesh}. (a) Geometry of mesh and listing of $\eta_i^{*}$ for all $\mathbf{V}_i$ in decreasing order. Surface $\Omega$ controlled by actuators.  (b) Configuration $\mathbf{V}_1$ associated with $\eta_1^{*}=\eta_{\text{max}}^{*}$. (c) Configuration $\mathbf{V}_2$ associated with $\eta_2^{*}$.  (d) Configuration $\mathbf{V}_{n-1}$ associated with $\eta_{n-1}^{*}$.  (e) Configuration $\mathbf{V}_n$ associated with $\eta_n^{*}=\eta_{\text{min}}^{*}$. }
\label{fig: eta_samples_spool}
\end{figure*}

We demonstrate this principle in Figs. \ref{fig: eta_samples_rectangle}-\ref{fig: eta_samples_spool}.  The FEM meshes displayed in Fig. \ref{fig: eta_samples_rectangle}a and Fig. \ref{fig: eta_samples_spool}a were converted into Case 1 state space models and expressed in discrete time with a zero order hold procedure.  All $\eta_i^{*}$ of \eqref{eq: eta_bound} were computed for each system using a value of $K=1000$ time steps and arranged in order of decreasing magnitude as displayed in Fig. \ref{fig: eta_samples_rectangle}a and Fig. \ref{fig: eta_samples_spool}a.  Fig. \ref{fig: eta_samples_rectangle}b-\ref{fig: eta_samples_rectangle}c and Fig. \ref{fig: eta_samples_spool}b-\ref{fig: eta_samples_spool}c display the temperature configurations associated with the first two (largest) values of $\eta_i^{*}$.  These are the temperature states which the systems ``preferentially'' drive towards and which require the smallest control energy to reach.  Recall that the control energy is applied to surface $\Omega$ of the builds, which in Fig. \ref{fig: eta_samples_rectangle}-\ref{fig: eta_samples_spool} is the uppermost line of nodes in the build geometries.  Qualitatively we observe that the temperature configurations which require the least control effort feature low temperatures away from $\Omega$ and broadly-heated regions at and around $\Omega$.  This result matches with the qualitative expectation that the ``easiest'' region of the build to maintain at elevated temperature is that in the immediate vicinity of the heat input $\mathbf{u}$.  Conversely, Fig. \ref{fig: eta_samples_rectangle}d-\ref{fig: eta_samples_rectangle}e and Fig. \ref{fig: eta_samples_spool}d-\ref{fig: eta_samples_spool}e display the temperature configurations associated with the last two (smallest) values of $\eta_i^{*}$.  These are the temperature configurations which the systems preferentially avoid and which require the largest control energy to reach.  We observe that these configurations are characterized by rapidly-alternating hot and cold regions, oftentimes far within the build interior while the exposed surface $\Omega$ must stay cool.  This result matches the qualitative expectation that it is ``difficult'' to precisely control internal temperatures far away from heat inputs placed on $\Omega$, especially under the constraint that $\Omega$ remain at relatively low temperature.     

\subsection{Required reconstruction/observation energy, case study}

\begin{figure*} [!tb]
\centering
\includegraphics[width=0.98\textwidth]{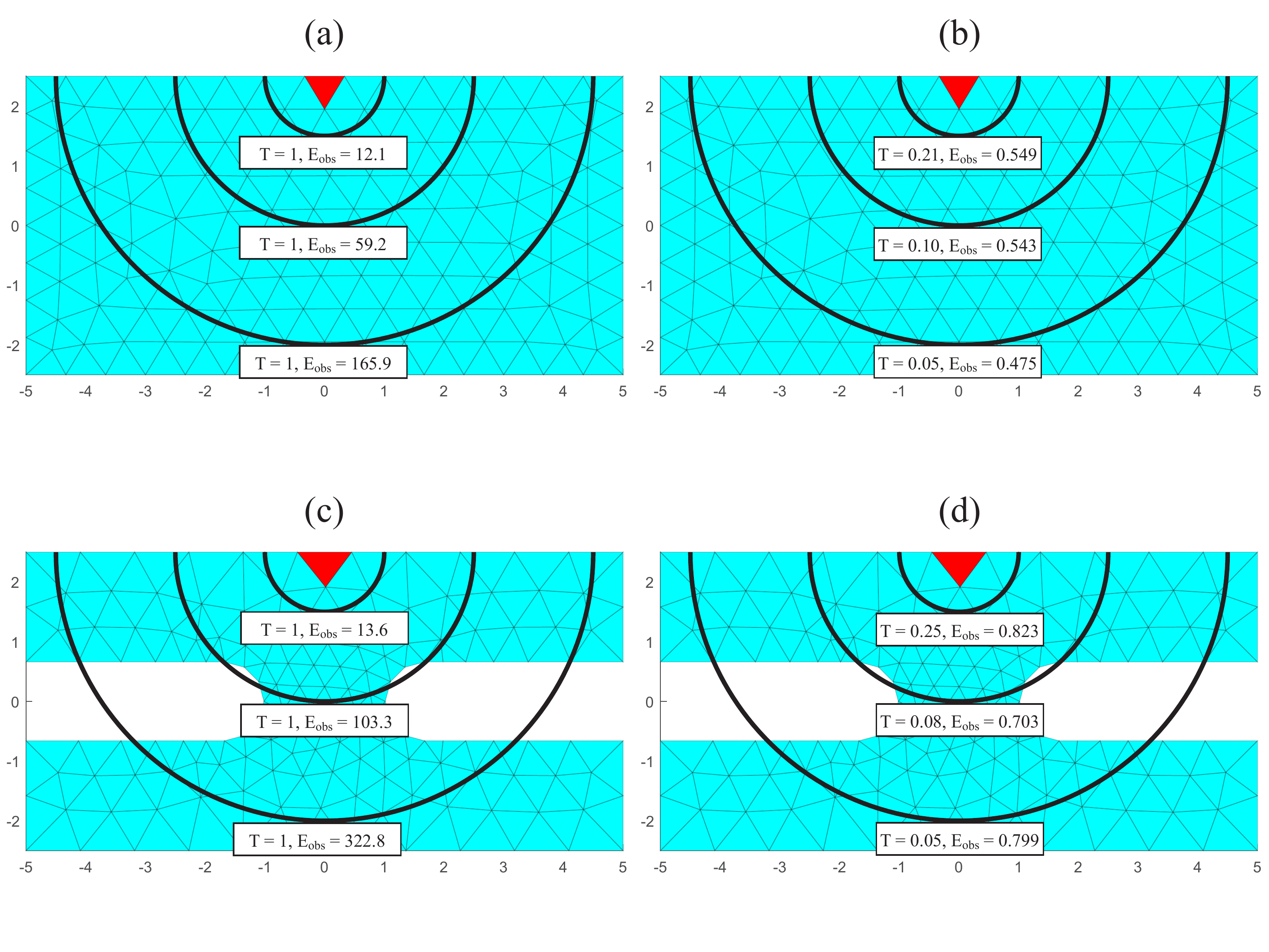}
\caption{\textbf{Absorbed sensor energy required to observe given $\mathbf{x}_f$ for build geometries of Figs. \ref{fig: eta_samples_rectangle}-\ref{fig: eta_samples_spool}}. Geometric details of test builds given in Figs. \ref{fig: eta_samples_rectangle}(a), \ref{fig: eta_samples_spool}(a).  Sensors for both geometries are observing nodes on $\Omega$ within element shaded in red. $\mathbf{x}_f$ defined by maintaining all nodes within regions enclosed by black semicircles at the temperatures indicated by $T$ and leaving non-enclosed nodes at $T=0$.  (a) Rectangular mesh, $\mathbf{x}_f$ defined by nodes at constant $T=1$.  (b) Rectangular mesh, $T$ of $\mathbf{x}_f$ defined such that $||\mathbf{x}_f||_2=1$.  (c) Spool-shaped mesh, $\mathbf{x}_f$ defined by nodes at constant $T=1$.  (d) Spool-shaped mesh, $T$ of $\mathbf{x}_f$ defined such that $||\mathbf{x}_f||_2=1$.     }
\label{fig: observability_energy}
\end{figure*}

We now study $E_{obs}$.  \eqref{eq: min_observe_energy} does not require inverting $\mathbf{W}_o$ therefore we apply this equation directly for the system geometries in the previous subsection.  Fig. \ref{fig: observability_energy} shows our case study setup and results.  Fig. \ref{fig: observability_energy}a and Fig. \ref{fig: observability_energy}b showcase the build geometry of Fig. \ref{fig: eta_samples_rectangle}, and Fig. \ref{fig: observability_energy}c and Fig. \ref{fig: observability_energy}d do the same for the build geometry of Fig. \ref{fig: eta_samples_spool}. The systems of Fig. \ref{fig: observability_energy} differ from those of Figs. \ref{fig: eta_samples_rectangle}-\ref{fig: eta_samples_spool} in that the nodes being observed are not all nodes on $\Omega$ but instead those nodes on $\Omega$ which exist in the element shaded in red.  We apply \eqref{eq: min_observe_energy} by constructing $\mathbf{x}_f[K]$ in two different ways: Fig. \ref{fig: observability_energy}a and Fig. \ref{fig: observability_energy}c maintain the nodes enclosed by their respective black semicircles at constant temperature $T=1$.  Fig. \ref{fig: observability_energy}b and Fig. \ref{fig: observability_energy}d maintain the nodes enclosed by their respective black semicircles at a $T$ such that $||\mathbf{x}_f[K]||_2=1$.  In both cases, all non-enclosed nodes of the system were left at $T=0$ (ambient temperature in the chosen units).  \par

We observe in Fig. \ref{fig: observability_energy}a and Fig. \ref{fig: observability_energy}c that as the radius of enclosed nodes maintained at $T=1$ increases so too does $E_{obs}$.  This agrees with the intuition that the ``difficulty'' of inferring information of subsurface dynamics from measurements of surface dynamics increases as these subsurface dynamics grow more extensive, as measured by the growth of $||\mathbf{x}_f[K]||_2^2$ relative to a constant $||\mathbf{y}[k]||_2^2$.  The system sensors must absorb increasing amounts of energy to reconstruct increasingly large amounts of information relative to the measured dynamics.  Build geometry plays a role in this information flow.  The spool-shaped geometry of Fig. \ref{fig: observability_energy}c cannot conduct heat to $\Omega$ from the interior as efficiently as the rectangular-shaped geometry of Fig. \ref{fig: observability_energy}a and therefore affords less efficient transmittance of energy (information) from the interior to the system sensors.  Accordingly $E_{obs}$ increases faster as the enclosed node radius increases in Fig. \ref{fig: observability_energy}c than in Fig. \ref{fig: observability_energy}a.

We observe the opposite trend in Fig. \ref{fig: observability_energy}b and Fig. \ref{fig: observability_energy}d.  As the radius of enclosed nodes increases, generally $E_{obs}$ decreases.  This change is due to the requirement that $||\mathbf{x}_f||_2=1$, which forces the maintained temperature $T$ of the enclosed nodes to decrease as the number of enclosed nodes increases.   A decrease in constant $T$ results in a decrease in measured $||\mathbf{y}[k]||_2^2$ for all $k$.  It follows from \eqref{eq: min_observe_energy} that a decrease in $||\mathbf{y}[k]||_2^2$ results in a decrease in $E_{obs}$. 

\section{Discussion and Conclusions}\label{sec:Con}

This paper demonstrated an initial exploration into a controls-based approach to estimating the internal temperature fields of parts being manufactured via the PBF process.  It was shown that an FEM-based linearization of the governing powder bed fusion physics produces a model that is unconditionally asymptotically stable, stabilizable, and detectable.  Four linearizations of the model were proposed, with each corresponding to a different set of available inputs and measurements of the PBF system.  Each of these linearized models is both structurally controllable and observable despite natural uncertainty in the governing system parameters.  We showed that having all-distinct eigenvalues is a sufficient condition for controllability/observability for time-invariant linearized models.  Furthermore, we present an initial characterization of the relative energy demands of controlling and observing time-invariant linearized models to better understand the practicality of system control and observation. 

The demonstration of controllability and observability affirms the feasibility of using state estimator concepts to acquire nodal temperature fields based only on observations of exposed surfaces and measured input signals.  The major benefit of this approach is the integration of the computational speed of LTI models with the enhanced prediction accuracy of closed-loop control.  This approach avoids the computational and financial burden of searching for appropriate process parameters via complex process models or design of experiment procedures.  Therefore these results present a step forward in realizing the goal of flexible, quasi-real time monitoring of the PBF process.

The controllability/observability results of this paper were given in the context of PBF thermal physics. These results easily abstract themselves from the context of this paper.  The discretization of arbitrary process physics into energy flow between a collection of nodes and edges will share the same controllability/observability properties shown here, provided that the discretized process model has a structure similar to ours.  We therefore anticipate that our controls theoretic approach to estimating scalar fields throughout a domain will be applicable in a variety of contexts.

We intend to pursue several avenues of research in light of of this result.  First, we intend to validate the performance of state estimators applied to models constructed using the procedure discussed here.  We will first conduct tests in simulation to assess basic questions of state estimator architecture and performance before conducting physical tests with simple test coupons.  These tests will be the subject of future papers on this subject.  We intend to investigate conditions under which time-varying linearized models of PBF are controllable/observable in the sense of Section \ref{subsec: classial_cont_obs}, and to explore the questions of controllability and observability for PBF actuation/measurement modes in which the system input/output relationships are nonlinear. 
 We also intend to explore the consequences of implementing \emph{adaptive meshing} on the stability/controllability/observability properties of the system.  An adaptive mesh is coarse except for a region of high node density which ``follows'' the heat source, thus allowing for much lower node counts and therefore lower computational burden.  This strategy complicates the model by making the network topology time-varying, however it allows for the modeling of more realistic aspects of the PBF process like material addition to the build.  We anticipate that this research will present a substantial amount of progress towards realizing in-situ, model-based process monitoring and control of PBF.
\section{Acknowledgements}
Financial support was provided by the member organizations of the Smart Vehicle Concepts Center, a Phase III National Science Foundation Industry-University Cooperative Research Center (www.SmartVehicleCenter.org) under grant NSF IIP 1738723.  The authors acknowledge technical support from ANSYS.
\begin{appendices}
\section{FEM-based discretization of PBF model}\label{Appendix: FEM_derivation}
The following derivation of the FEM discretization of our model follows the presentation of FEM heat transfer given in \cite{Cook89}.  This subsection is meant to be a brief introduction to the principles of FEM, readers interested in a complete description of FEM theory and practice should consult \cite{Cook89}. \par 

The FEM discretizes the problem domain into a series of elements which are bounded by nodes.  Fig. \ref{fig: PBF_BCs}b demonstrates such a discretization.  In this manner \eqref{eq:fourier_weak}, which describes the energy flow throughout the domain, is divided into a summation of functionals $\Pi_e$ which describe the energy flow within each element:

\begin{equation}
\Pi=\sum_e\Pi_e
=\sum_e \int_{V_e} \left( \frac{1}{2}(\nabla T)^\prime \mathbf{\kappa}\nabla T  + \rho c\dot{T}T \right)dV
-\int_{S_e}\left(u_B T\right) dS.\label{eq: fourier_weak_elemental}
\end{equation}

\noindent The subscripted terms $V_e$ and $S_e$ in \eqref{eq: fourier_weak_elemental} denote the volume and boundary of the $e^{th}$ element, respectively. $u_B$ represents the applied heat flux along the boundary of $V_e$.  \par 

The continuous temperature field $T$ is approximated in space within each element by means of interpolation.  Assume that the $e^{th}$ element is bounded by $n_e$ nodes.  Let $\mathbf{x_e}(t)\in\mathbb{R}^{n_e}$ collect the temperatures at these nodes.  Since the nodes bounding the $e^{th}$ element are singular points in space, the temperatures at these points, $\mathbf{x_e}(t)$, depend only on time.  The interpolation of $T$ is performed with a series of \emph{shape functions}, which are collected in a vector $\mathbf{N}_e(\mathbf{v})=[N_{1,e}(\mathbf{v}),\dots,N_{n_e,e}(\mathbf{v})]\in\mathbb{R}^{1\times n_e}$.  These shape functions are designed to enforce continuity in the approximated temperature field with neighboring elements, and construct the interpolation:

\begin{equation}
T(\mathbf{v},t)=\mathbf{N}_e(\mathbf{v})\mathbf{x_e}(t),\textnormal{ }\mathbf{v}\in V_e.\label{eq: ShapeFn}
\end{equation}

\eqref{eq: ShapeFn} is substituted into \eqref{eq: fourier_weak_elemental}, which reduces the functional to the form shown in \eqref{eq: FEM_weak_form}.

\begin{equation}
\begin{split}
&\Pi=\sum_e\frac{1}{2}\mathbf{x_e}^\prime\mathbf{K}_e\mathbf{x_e}+\mathbf{x_e}^\prime\mathbf{M}_e\dot{\mathbf{x}}_\mathbf{e}-\mathbf{x_e}^\prime\mathbf{R}_e\\
&\mathbf{K}_e=\int_{V_e}\mathbf{B}_e^\prime\mathbf{\kappa}\mathbf{B}_edV_e\\
&\mathbf{M}_e=\int_{V_e}\mathbf{N}_e^\prime\rho c\mathbf{N}_edV_e\\
&\mathbf{R}_e(t)=\int_{S_e}\mathbf{N}_e^\prime u(\mathbf{v},t)dS_e\\
&\mathbf{B}_e=\begin{bmatrix}
\left[\frac{\partial\mathbf{N}_e}{\partial x}\right]^\prime &\left[\frac{\partial\mathbf{N}_e}{\partial y}\right]^\prime &\left[\frac{\partial\mathbf{N}_e}{\partial z}\right]^\prime 
\end{bmatrix}^\prime. 
\end{split}\label{eq: FEM_weak_form}
\end{equation}

Note that in \eqref{eq: FEM_weak_form}, the boundary term $u_B$ of \eqref{eq: fourier_weak_elemental}, which encompasses both the heat flux function $u(\mathbf{v},t)$ on $\Omega$ and the Dirichlet boundary on $\Lambda$, was replaced by $u(\mathbf{v},t)$.  We will incorporate the Dirichlet boundary after the equations have been formulated for all nodes.  The elemental functionals of \eqref{eq: FEM_weak_form} are \emph{assembled} by describing their constituent element node numbers and $\mathbf{K}_e$, $\mathbf{M}_e$, and $\mathbf{R}_e(t)$ indices in terms of a global node ordering.  This procedure is demonstrated in Fig. \ref{fig:Kbuild}.  We denote $\mathbf{x}_{\Pi}$, $\mathbf{K}_{\Pi}$, $\mathbf{M}_{\Pi}$, and $\mathbf{R}_{\Pi}(t)$ as the \emph{global} (assembled) counterparts to the quantities of \eqref{eq: FEM_weak_form}.  Here, $\mathbf{x}_{\Pi}\in\mathbb{R}^n$ collects the temperatures of all $n$ nodes in the FEM mesh, including those nodes with constrained temperatures due to laying on $\Lambda$, $\mathbf{K}_{\Pi}\in\mathbb{R}^{n\times n}$ describes the conductivity between all nodes of the mesh, $\mathbf{M}_{\Pi}\in\mathbb{R}^{n\times n}$ describes the thermal mass between all nodes of the mesh, and $\mathbf{R}_{\Pi}(t)\in\mathbb{R}^n$ distributes the load $u(\mathbf{v},t)$ among all nodes of the mesh:   

\begin{figure} [!tb]
\centering
\includegraphics[width=0.75\textwidth]{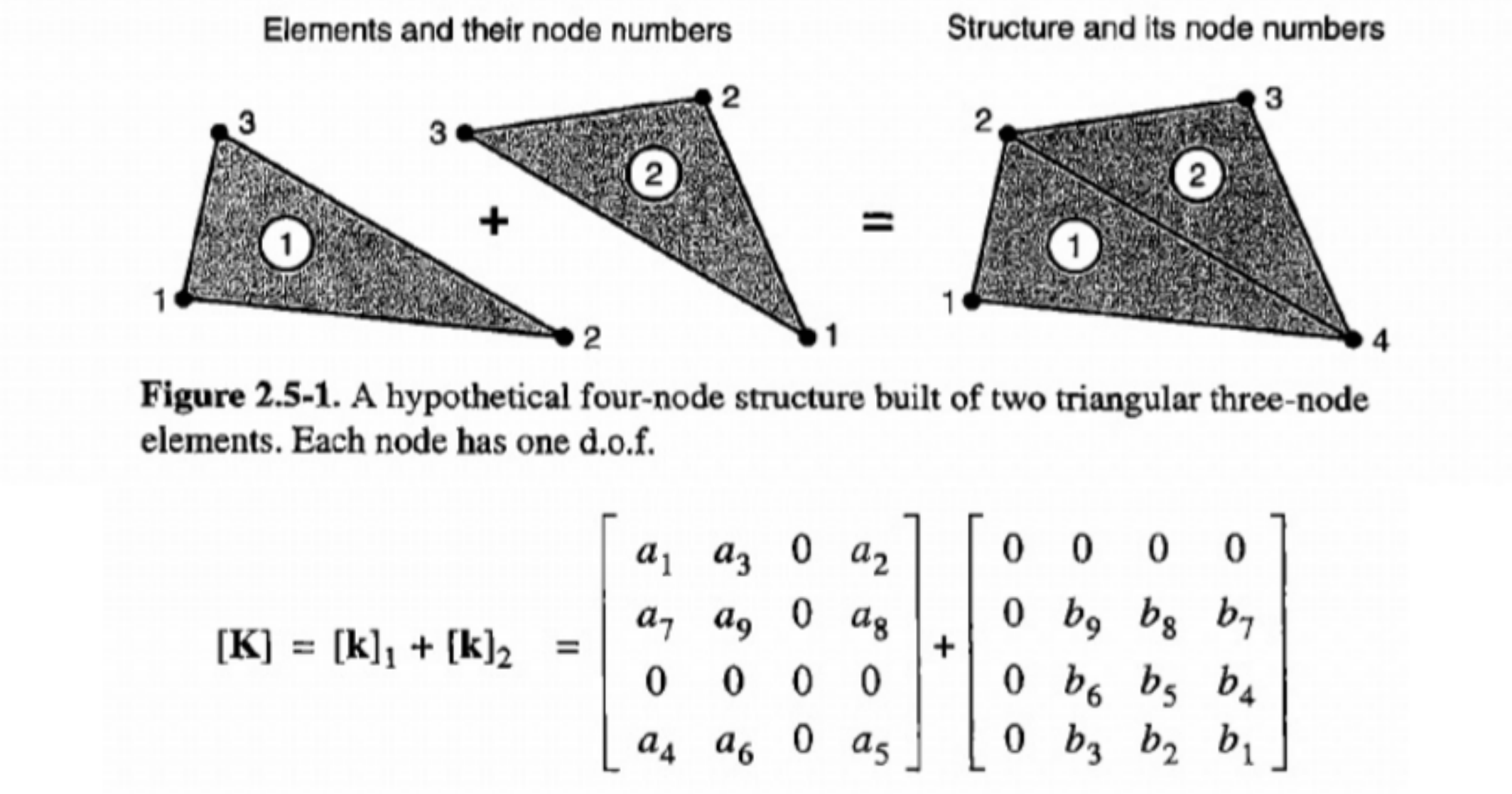}
\caption{Global system construction process.  Reproduced from \cite{Cook2002} with permission from John Wiley and Sons, Inc.}
\label{fig:Kbuild}
\end{figure}    

\begin{equation}
\begin{split}
&\Pi=\frac{1}{2}\mathbf{x}_{\Pi}^\prime \mathbf{K}_{\Pi}\mathbf{x}_{\Pi}+\mathbf{x}_{\Pi}^\prime \mathbf{M}_{\Pi}\dot{\mathbf{x}}_{\Pi}-\mathbf{x}^\prime \mathbf{R}_{\Pi}\\
&\mathbf{K}_{\Pi}=\sum_e\int_{V_e}\mathbf{B}_e^\prime\mathbf{\kappa}\mathbf{B}_edV\\
&\mathbf{M}_{\Pi}=\sum_e\int_{V_e}\mathbf{N}_e^\prime\rho c\mathbf{N}_edV\\
&\mathbf{R}_{\Pi}(t)=\sum_e\int_{S_e}\mathbf{N}_e^\prime u(\mathbf{v},t)dS\\
&\mathbf{B}_e=\begin{bmatrix}
\left[\frac{\partial\mathbf{N}_e}{\partial x}\right]^\prime &\left[\frac{\partial\mathbf{N}_e}{\partial y}\right]^\prime &\left[\frac{\partial\mathbf{N}_e}{\partial z}\right]^\prime 
\end{bmatrix}^\prime.
\end{split}\label{eq: FEM_weak_form_global}
\end{equation}

The functional given in \eqref{eq: FEM_weak_form_global} describes the energy flow through the system for any temperature field within $V$, as approximated by \eqref{eq: ShapeFn}.  The well-known \emph{Principle of Stationary Energy} states that the system will arrive at a temperature field $\mathbf{x}$ which makes $\Pi$ stationary, ie $d\Pi=0$ for small changes in the approximated temperature field $d\mathbf{x}_{\Pi}$.  This stationary point corresponds to a minimum value of $\Pi$.  Qualitatively, this means that the ``disturbance'' within the system -- the amount of energy dissipated and stored within it as a response to the load $u(\mathbf{v},t)$ -- is made as small as possible relative to the energy delivered into it through the load at every instant in time.  We write $d\Pi$ as a differential:

\begin{equation}
d\Pi=\frac{\partial\Pi}{\partial[\mathbf{x}_{\Pi}]_1}d[\mathbf{x}_{\Pi}]_1+\frac{\partial\Pi}{\partial[\mathbf{x}_{\Pi}]_2}d[\mathbf{x}_{\Pi}]_2+\cdots+\frac{\partial\Pi}{\partial[\mathbf{x}_{\Pi}]_n}d[\mathbf{x}_{\Pi}]_n=0,\nonumber
\end{equation}

\noindent Here, $[\mathbf{x}_{\Pi}]_i$ denotes the $i^{th}$ element of $\mathbf{x}_{\Pi}$.  It is clear that $d\Pi=0$ for \emph{any} combination of nonzero nodal temperature variations $d[\mathbf{x}_{\Pi}]_i$ in the mesh if and only if $\frac{\partial\Pi}{\partial[\mathbf{x}_{\Pi}]_i}=0$, $\forall i=\{1,2,\dots,n\}$, denoted as $\frac{\partial\Pi}{\partial\mathbf{x}_{\Pi}}=0$.  Applying this criteria to \eqref{eq: FEM_weak_form_global} constructs a set of $n$ coupled ordinary differential equations, which are expressed in matrix form in \eqref{eq: FEM_solution}. 

\begin{equation}
\begin{split}
&\mathbf{M}_{\Pi}\dot{\mathbf{x}}_{\Pi}+\mathbf{K}_{\Pi}\mathbf{x}_{\Pi}=\mathbf{R}_{\Pi}(t).\\
\end{split}\label{eq: FEM_solution}
\end{equation}

\eqref{eq: FEM_solution} is then \textit{reduced}.  This procedure treats the constrained nodes on $\Lambda$ as sources of constant heat flux on the nodes immediately adjacent to $\Lambda$ and in doing so removes the nodes on $\Lambda$ from the system.  This procedure is best taught by case study, and FEM textbooks such as \cite{Cook89} walk the reader through illustrative examples.  Following this procedure constructs the system of ODEs $\mathbf{M}\dot{\mathbf{x}}+\mathbf{Kx}=\mathbf{R}(t)$, which completes the derivation of $\mathbf{M}$, $\mathbf{K}$, and $\mathbf{R}(t)$ of \eqref{eq: FEM_ODEs}.

\section{Derivation of $\mathbf{Bu}$ from $\mathbf{r}(t,\mathbf{u})$}\label{Appendix: Bu_derivation}

We suppose that $u(\mathbf{\bar{v}},t)$ is arbitrary as in Case 1.  Doing so results in $\mathbf{r}(t,u)$ assuming the structure shown in \eqref{eq: Input_A.A}, as shown in Appendix \ref{Appendix: FEM_derivation}.  Here, $dS_e$ represents integration of $u(\mathbf{v},t)\mathbf{N}_e^\prime$ over all faces of the $e^{th}$ element in the FEM mesh.  $u(\mathbf{v},t)=u(\bar{\mathbf{v}},t)$ when $\mathbf{v}\in\Omega$ and 0 otherwise. The summation over $e$ represents the assembly of all such \emph{elemental} integral-defined vectors as defined in Appendix \ref{Appendix: FEM_derivation}.  
\begin{equation}
    \mathbf{r}(t,\mathbf{u})=\mathbf{M}^{-1}\mathbf{R}=\mathbf{M}^{-1}\sum_e\int_{S_e}u(\mathbf{v},t)\mathbf{N}_e^\prime dS_e.\label{eq: Input_A.A}
\end{equation}
The arbitrary nature of $u(\mathbf{\bar{v}},t)$ under Actuation mode \textbf{A.A} allows for independent control of the heat flux over all elements on $\Omega$, which we use to linearize  \eqref{eq: Input_A.A} via a quantization procedure as described pictorially in  Fig. \ref{fig: GaussianDiscretization}.  We construct an FEM mesh such that the element faces on $\Omega$ are sufficiently small to justify approximating the intensity of $u(\mathbf{v},t)$ overtop them as constant.  The assumed-spatially constant laser intensity over the $e^{th}$ element is denoted as $u_e(t)$.  Under this assumption, the summand of \eqref{eq: Input_A.A}, $\mathbf{R}_e(t)$, takes the form:

\begin{sidewaysfigure*} [!tb]
\centering
\includegraphics[width=1.0\textwidth]{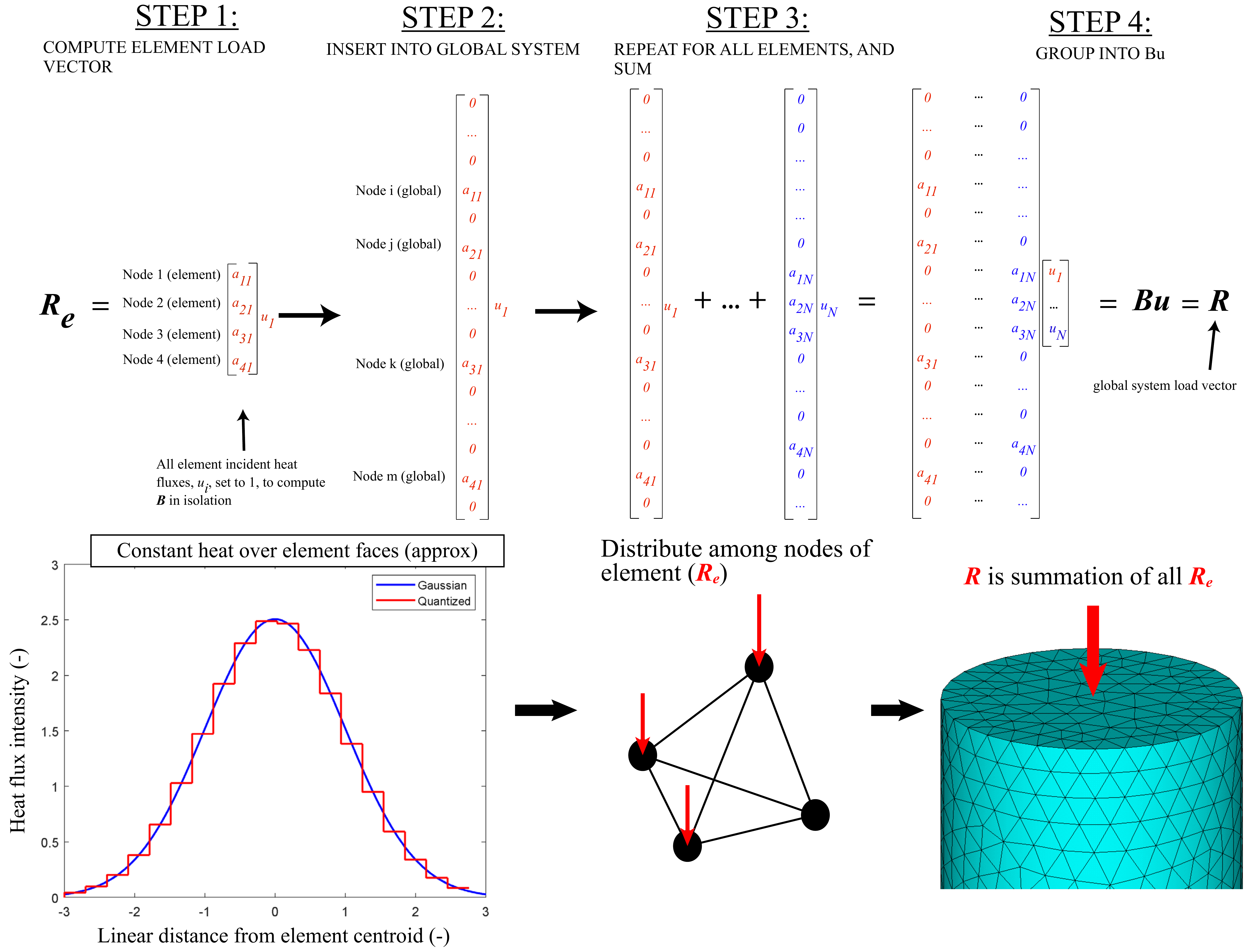}
\caption{Workflow for linearizing system input by quantizing heat source intensity as uniform over the faces of the elements in the FEM mesh.  Reasonable approximation of the load distribution with this method requires element size to be much smaller than the laser beam diameter.}
\label{fig: GaussianDiscretization}
\end{sidewaysfigure*} 

\begin{equation}
\mathbf{R}_e(t)=\int_{S_e}u_e(t)\mathbf{N}_e^\prime dS_e=u_e(t)\int_{S_e}\mathbf{N}_e^\prime dS_e.\label{eq: ElementLoadVector}
\end{equation}

\noindent It is assumed that the FEM mesh remains static with respect to time, such that all nodes remain fixed in space.  Therefore the shape functions $\mathbf{N}_e$ and element geometry $S_e$ are assumed constant with respect to time and therefore the integral $\int_{S_e}\mathbf{N}_e^\prime dS$ produces a constant vector of weights, scaled by $u_e(t)$ as a proportionality constant. As Fig. \ref{fig: GaussianDiscretization} demonstrates for a hypothetical 4-node element, the summation $\sum_e\mathbf{R}_e$ can thus be broken into a matrix product $\mathbf{B}\mathbf{u}(t)$.  $\mathbf{B}$ collects these integral-computed weighting vectors in the global system formulation, and $\mathbf{u}$ collects the assumed-uniform heat inputs over all elements. The control input $\mathbf{u}(t)$ is chosen to be the time-varying intensity of the Gaussian distribution at the centroid of all elements.  Elements without faces belonging to $\Omega$ are assigned a value of $u_e(t)=0$ for all time. As such, $\mathbf{r}(t,\mathbf{u})=\mathbf{M}^{-1}\mathbf{R}(t)$ may be expressed as $\mathbf{r}(t,\mathbf{u})=\mathbf{M}^{-1}\mathbf{B}_0\mathbf{u}(t)$.  Redefining $\mathbf{B}=\mathbf{M}^{-1}\mathbf{B}_0$ we arrive at the expression $\mathbf{r}(t,\mathbf{u})=\mathbf{Bu}(t)$. 
\end{appendices}


\bibliography{biblio}
\bibliographystyle{IEEEtran}
\end{document}